\documentclass[twocolumn]{revtex4-1}
\usepackage{graphicx}
\usepackage{epstopdf}
\graphicspath{{./figures/}}
\usepackage{amssymb}
\usepackage{float}
\usepackage{physics}
\usepackage{commath} 
\usepackage{amsmath}
\usepackage[table,xcdraw]{xcolor} 
\usepackage{multirow}
\usepackage{hhline}
\usepackage{tabularx}

\begin{document}
\title{Improving Qubit Readout with Hidden Markov Models}
\author{Luis A. Martinez}
\author{Yaniv J. Rosen}
\author{Jonathan L. DuBois}
\date{May 29, 2020}

\begin{abstract}
We demonstrate the application of pattern recognition algorithms via hidden Markov models (HMM) for qubit readout. This scheme provides a state-path trajectory approach capable of detecting qubit state transitions and makes for a robust classification scheme with higher starting state assignment fidelity than when compared to a multivariate Gaussian (MVG) or a support vector machine (SVM) scheme. Therefore, the method also eliminates the qubit-dependent readout time optimization requirement in current schemes. Using a HMM state discriminator we estimate fidelities reaching the ideal limit. Unsupervised learning gives access to transition matrix, priors, and IQ distributions, providing a toolbox for studying qubit state dynamics during strong projective readout. 
\end{abstract}

\maketitle

\section{Introduction}
Quantum processors employing superconducting qubits are now reaching new milestones in their simulation \cite{barends2015digital,wendin2017quantum, colless2018computation,yan2019strongly} and computational capabilities \cite{arute2019quantum}. There are numerous technical challenges in the implementation of a fault tolerant quantum processor, but at the core is the ability to generate high fidelity gates \cite{lucero2008high, rol2017restless}, perform quantum error correction \cite{reed2012realization,corcoles2015demonstration}, and the ability to make high-fidelity qubit readout measurements \cite{walter2017rapid}. In particular, high-fidelity single shot qubit readout enables faster quantum protocols while simultaneously allowing for reduced errors in their characterization. Apart from improving $T_1$ times of superconducting qubits \cite{nersisyan2019manufacturing,place2020new}, optimizing hardware design and configuration \cite{walter2017rapid}, and invoking new qubit-cavity coupling schemes \cite{dassonneville2020fast}, readout fidelity may be improved by applying classification schemes utilizing machine learning algorithms \cite{magesan2015machine,seif2018machine}. 

Here we demonstrate the application of pattern recognition algorithms, via hidden Markov models \cite{rabiner1986introduction}, to the heterodyned readout signal of a superconducting qubit \cite{blais2004cavity}. The Markov structure allows for a state-path trajectory approach by discretizing each shot into a sequence of uncorrelated segments. The result is a robust starting state classification scheme with higher fidelity than when compared with multivariate Gaussian (MVG) and support vector machines (SVM) classifiers \cite{cortes1995support}. The advantage arises from the ability to detect transitions with high probability and, thus, circumvent measurement obfuscation caused by qubit state relaxation. In addition, the application of hidden Markov models for qubit readout can naturally be extended to multi-level qudit systems. Unsupervised learning with hidden Markov models provide the capability of extracting distribution parameters, transition matrices, and starting state probabilities (priors), therefore, providing a valuable toolbox for qubit readout and measurement error correction \cite{sun2018efficient,geller2020rigorous}. 

This paper is organized as follows. First, an example illustrating the evolution of the readout signal in the IQ plane is presented, followed by a description of the experimental system used to generate the experimental data. We continue with a brief description of the MVG and SVM classifiers and define the fidelity metrics before detailing the implementation of the hidden Markov model (HMM) classifier. Next, we extract the statistical variations associated with training HMMs and calculate classification errors. Finally, we calculate the readout fidelity of a HMM state classifier and compare it with the ideal fidelity metric defined in reference \cite{magesan2015machine}.

\begin{figure}[h]
   \centering
   \includegraphics[width  = .35\textwidth]{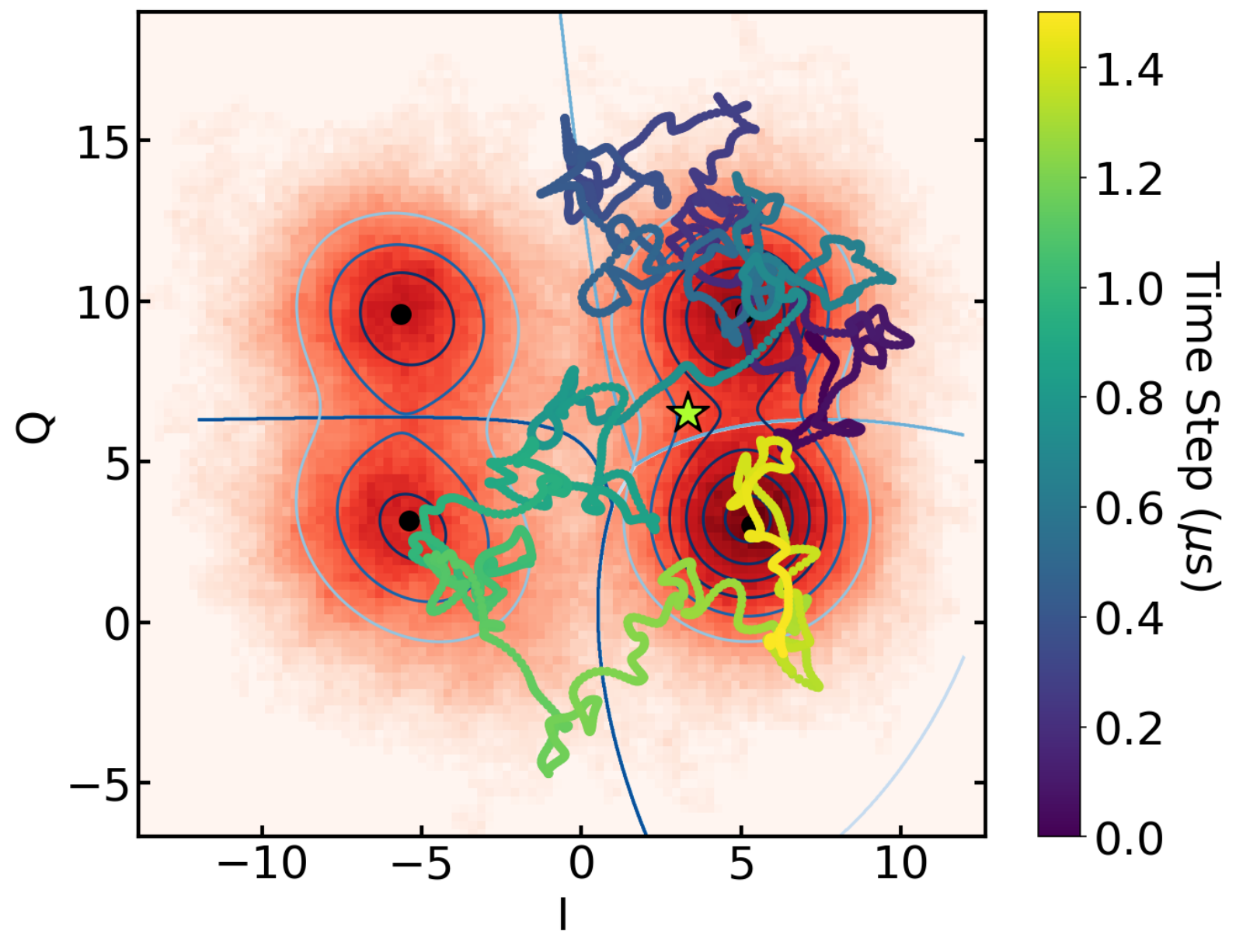}   
    \caption{Running average (colored line) of the heterodyned signal of a single shot in a two-qubit four-state system. The different colors correspond to time evolution. The star marker denotes the demodulated IQ value over the entire measurement time. Bayes classifier and contour lines representing probability distributions shown for reference.}
   \label{fig:trajectory}
\end{figure}

Random noise and qubit decay processes reduce readout fidelity. In Fig. \ref{fig:trajectory} the trajectory of a single shot measurement for two coupled qubits in a 3D CQED system \cite{wu2020high} is tracked in the IQ plane. For reference, Fig. \ref{fig:trajectory} also includes a Bayes classifier trained on several single shots for each qubit state. The contour lines represent the probability distributions learned with a general mixture model. From the running average of the heterodyned signal (colored line) we see the signal starts near the prepared $\ket{0,1}$ state (purple), wanders around the IQ plane, and finally decays to the ground state $\ket {0,0}$ (yellow). Integration over the total readout time, denoted by the star marker, illustrates that this shot would had been classified to state $\ket{0,1}$ with low probability. This example illustrates that, apart from optimizing hardware parameters \cite{walter2017rapid}, choosing an appropriate readout integration time plays an important role in mitigating qubit relaxation.

\begin{figure*}[t]
   \centering
   \includegraphics[width = 1\textwidth]{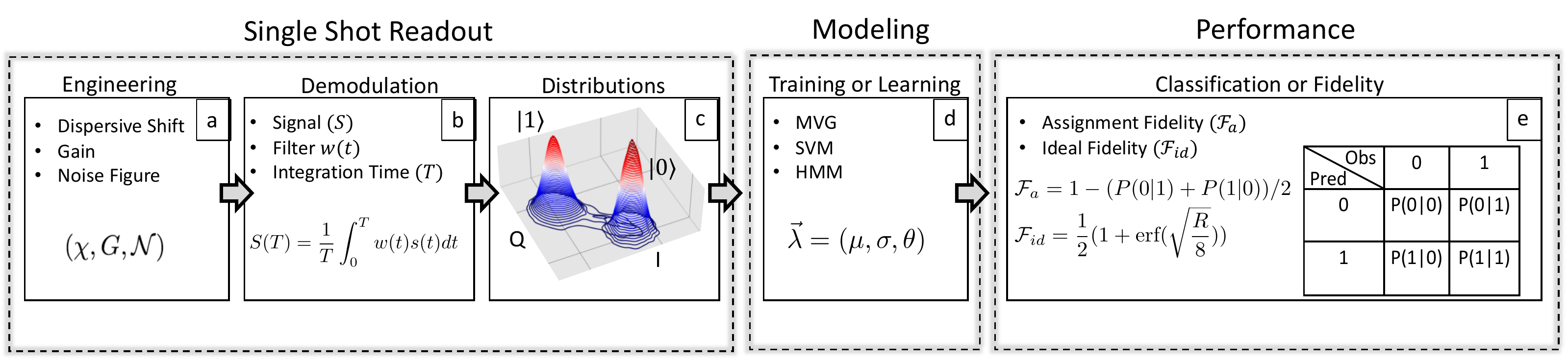}   
    \caption{Summary of Qubit Readout. (a) Factors determining readout fidelity begin with the quantum hardware; Hamiltonian parameters, and SNR. (b) Apart from hardware, fidelity may be improved at the demodulation stage. Conventional demodulation scheme requires tuning to an optimal integration time in which the distributions are approximately Gaussian (c). (d) Regression analysis can be used to train models for improved classification performance. (e) Classification errors are extracted from the misclassification probabilities.}
   \label{fig:schematic}
\end{figure*}

\section{Methods}
The experimental quantum platform is a 3D cavity QED \cite{blais2004cavity,wu2020high} system utilizing a strong-projective dispersive measurement scheme \cite{wallraff2005approaching}. In this platform the qubit state information is encoded in the amplitude and phase of the readout signal. For a single shot measurement, a readout pulse of width $W$ and radio frequency $\omega_r$ is applied to the readout resonator, filtered, and amplified. At this stage, as indicated in Fig. \ref{fig:schematic}(a), a fidelity limit is imposed by the signal to noise ratio (SNR) and the CQED system parameters \cite{walter2017rapid}. For the purpose of this work, these parameters are to be associated with hardware and, therefore, assumed to be fixed after some initial optimization. Next, the amplified signal is mixed down, with a RF-mixer and a local oscillator tone at frequency $\omega_{LO}$, to an intermediate frequency $\Omega_{IF} = \omega_r-\omega_{LO}$. The IF signal is then digitally decomposed in quadrature and demodulated by picking out the Fourier component at $\Omega_{IF}$. The demodulation process integrates both the in-phase and quadrature components for a total integration time $T_{int}$ (Fig. \ref{fig:schematic}(b)). At this stage the readout fidelity may be optimized by appropriately tuning of the integration time. Insufficient state distinguishability may arise for relatively short integration times, whereas qubit decay obfuscates the readout signal for long times. The demodulation of each single shot results is a $(I,Q)$ coordinate and the histogram of several shots forms the probability distribution in the IQ-plane as illustrated in Fig. \ref{fig:schematic}(c).

Preparing a state classifier involves a training procedure in which a training dataset is used to learn the model parameters (Fig. \ref{fig:schematic}(d)). For the multivariate Gaussian (MVG) model, the training solely consists of learning the mean array {$\vec \mu$} and covariance matrix $\mathbf{\Sigma}$. The goodness-of-fit of the MVG model is severely affected for longer readout times because the IQ distributions for states exhibiting state transitions become skewed and, therefore, non-Gaussian \cite{gambetta2007protocols}. Support vector machines (SVM) provide both supervised and unsupervised learning capabilities \cite{cortes1995support,ben2001support}. Because SVMs are geometric models and can, therefore, partially circumvent random noise processes, they provide excellent classification results by finding an optimal hyperplane which provides maximum separation between clusters. However, SVMs also succumb to $T_1$ effects, and so, an optimal integration time is required for maximum readout fidelity.

\subsection{Fidelity Metrics}
Before describing HMMs we define the fidelity metrics used herein, and remind the reader we are operating in the strong projective measurement limit.
In the absence of qubit state decay and assuming Gaussian noise, the IQ distributions for each readout state are Gaussian. The ideal fidelity, ${\cal F}_{id}$, defined by the misclassification probability, is computed from the integration of the overlapped regions of the projected Gaussian probability distributions \cite{magesan2015machine};
\begin{align}
{\cal F}_{id} = \frac{1}{2}(1 + \text{erf}(\sqrt{\frac{R}{8}})).
\label{eq:ideal_fidelity}
\end{align}
Here, R is a measure of the separation between the two distributions in question \cite{magesan2015machine}.
\begin{align}
R = \frac{(\langle S_0\rangle - \langle S_1\rangle)^2}{\text{var}(S)}
\label{eq:R_value}
\end{align}
$S$ denotes the measurement outcome after the integration of the signal, i.e. the demodulated value, and $\text{var}(S)$ is the variance of $S$. In practice, the IQ probability distributions are also well modeled by Gaussian distributions so long we operate in the limit where the integration time is much less than the relaxation time. However, for relatively long integration times ($T_{int} \gtrsim 5\%\times T_1$), the distributions are skewed by relaxation transitions and a Gaussian model is no longer adequate. 

If the readout state probability distributions are not Gaussian, the method of calculating fidelity described above in Eq. \eqref{eq:ideal_fidelity} is not suitable. Instead, fidelity of classification systems may generally be assessed through various statistical figures of merit \cite{reagor2018demonstration}. Here we use the assignment fidelity, (${\cal F}_a$), to compare the MVG, SVM, and HMM classifiers. The assignment fidelity is adapted from a more general distortion measure which is equivalent to the infidelity. Introduced by Shanon's information theory \cite{shannon1959coding}, it gives a quantitative measure for performance of classification systems which generalizes to the confusion matrix formalism illustrated in Fig. \ref{fig:schematic}(e). For the two state case the assignment fidelity is
\begin{equation}
{\cal F}_a = 1-\frac{1}{2}(P(0|1) + P(1|0)),
\label{eq:fidelity}
\end{equation}
where $P(i|j)$ is the probability the label $i$ is assigned when state $j$ is prepared. Note, that this definition makes $P(i|j)$ dependent on the fidelity of the gate used to prepare the state, and the starting state population at the start of the readout measurement.

\subsection{Hidden Markov Models for Qubit Readout}
Hidden Markov models are a special case of Bayesian networks in which underlying hidden stochastic Markov processes yield observations which themselves are associated with a probability distribution. HMMs have been used to measure the relaxation time of the nuclear spin state of a nitrogen-vacancy defect in diamond \cite{dreau2013single}, and more recently in a proposed robust readout scheme for bosonic systems in the dispersive coupling regime \cite{hann2018robust}. For this work the relevant parameters (see appendix section \ref{HMM_details}) of a HMM are \cite{rabiner1986introduction}:
\begin{align}
O & = \text{(I,Q) pair observation sequence}\\
D & = \{d\},~\text{set of qubit states} \nonumber \\
N & = \text{number of qubit states in the model} \nonumber \\
\pi & = \text{set of initial state distributions}\nonumber\\
B_d & = \text{(I,Q) pair probability distribution} \nonumber \\ &\quad~ \text{given state}~ d \in D \nonumber \\ 
\mathbf{A} & = [a_{ij}], \text{state transition matrix}\nonumber
\end{align}

Preparing a readout measurement as a Markov chain requires partitioning a single shot of total time $t$ into several uncorrelated segments. Note that the total readout time $t$ does not necessarily correspond to the readout pulse width $W$, and typically $t<W$. Each segment is demodulated for a short time interval $\Delta t$, resulting in a single observation in the form of an IQ pair; $O_i = (I_i,Q_i)$. Therefore, each shot becomes a discretized sequence of observations of size $n = t/\Delta t$, where $n$ is an integer. The emission probability distribution, $B_d\equiv P(O|d)$, is the probability that the observation pair $(I,Q)$ was emitted from state $d$. For qubit readout, $B_d$ corresponds to the 2-dimensional probability distributions in IQ-space that randomize the readout measurement based on the hidden state (e.g., Fig. \ref{fig:schematic}(c)). The hidden states are identified with the qubit states; for two states $D = \{{\ket 0}, {\ket 1}\}$. The Markov assumption requires that the state $d_i$ be only dependent on the preceding state $d_{i-1}$. This is satisfied since in this regime, the probability of transitioning from the excited state in one observation segment to the ground state in the next is fixed by $P_e(\Delta t) = e^{-\Delta t/T_1}$. 

The transition matrix ${\bf A}=[a_{ij}]$ gives the transition probabilities between the qubit states. The off-diagonal elements of the form $a_{ij}$ for $i>j$ are associated with qubit state relaxation, and elements in which $i < j$ can be attributed to qubit excitations, e.g. due to heating. The initial state distributions $\pi$ represent the initial state probability of state $d$, i.e. the priors. For example, in the ideal case in which the excitation rate is zero ($a_{01}=0$), the transition matrix (${\bf A}$) for the two state case is given by
\begin{equation}
a_{ij} = 
\begin{bmatrix}
 1 & 0 \\ 
 1-e^{-\Delta t/T_{1,eff}} & e^{-\Delta t/T_{1,eff}} 
   
\end{bmatrix},
\end{equation}
where $i,j \in \{0,1\}$, and $T_{1,eff}$ is the effective relaxation time which accounts for measurement induced dephasing \cite{boissonneault2009dispersive}. In practice, the heating rate is not necessarily zero and can be extracted from the learned transition matrix element, $\Gamma_{10} = a_{10} / \Delta t$.

\begin{figure}[t]
  \centering
   \includegraphics[width = .35\textwidth]{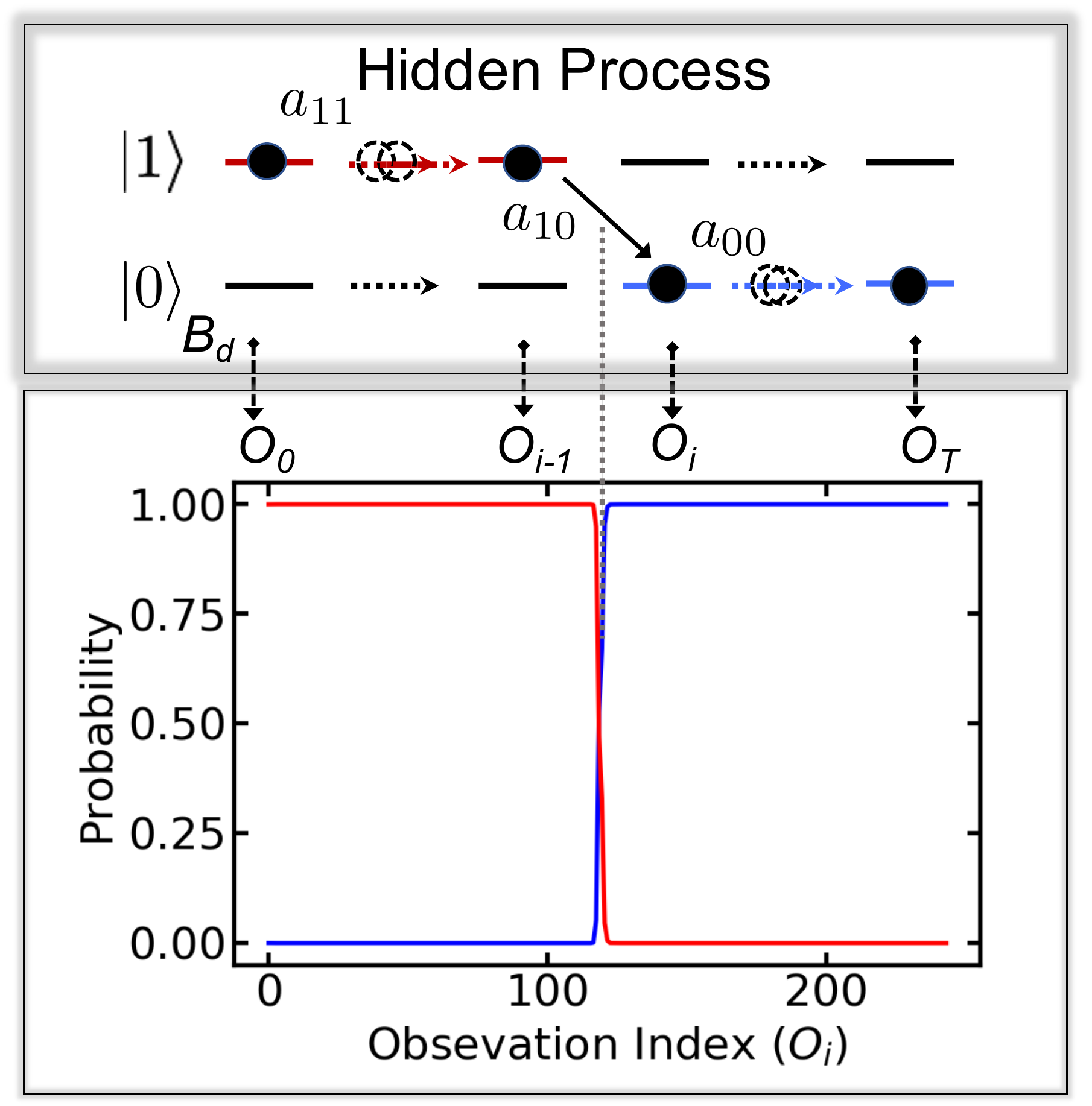}   
    \caption{State path prediction with hidden Markov models for readout of a single 20 $\mu$s shot. Each shot is discretized into a sequence of 80 ns observations ($O_i$), and each observation is an (I,Q) pair emitted from the hidden state having an emission probability distribution $B_d$. The forward-backward algorithm computes the probability of being in the ground (blue) or excited (red) state, $ \{{\ket 0}, {\ket 1}\}$ respectively. The probability the qubit was in the excited state at the start of the measurement was 99.98\%.}
   \label{fig:HMM}
\end{figure}

A HMM is defined by the set of parameters $\lambda = ({\bf A},  B,  \pi, N )$. There are three well known solved problems with hidden Markov models, and they are briefly re-summarized in the following \cite{rabiner1986introduction}. First, given a HMM $\lambda = ({\bf A}, B,  \pi, N )$ and a sequence of observations $O$ one can determine the probability of the sequence given the model $\lambda$, $P(O|\lambda)$. This probability can be computed in a straightforward fashion by summing the product of the observation sequence probability and the state sequence probability over all possible state sequences. However, this process is computationally intensive. In practice, either of the so-called forward or backward algorithms are used to compute $P(O|\lambda)$ \cite{rabiner1986introduction}. Note that both the forward and backward algorithms enable the efficient applicability of HMMs. Second, given a model $\lambda$ and observation sequence $O$, an optimal hidden state sequence can be computed. This feature of HMMs allows for the prediction of the optimal state sequence by calculating the probability of being in state $d$ at observation $O_i$. The predicted hidden state sequence is then composed by selecting the most probable state at each observation. And third, given an observation sequence $O$ and the number of hidden states $N$, the model parameters can be computed by solving the maximum likelihood problem using the Baum-Welch algorithm \cite{baum1970maximization}. This key feature enables the unsupervised training of HMMs. 

The application of the forward-backward algorithm via HMMs for single shot readout yields a predicted state path that is based on the probability of being in state $d$ at each observation $O_i$ of the sequence. Because the forward-backward algorithm finds the best-fit state sequence given an observation sequence and model, the result is the determination of the qubit state at each index with high probability. In particular,  the determination of the qubit state $d_0$ at the start of the readout measurement, can be made with high probability even in cases where a relaxation transition occurs during the readout measurement. This is of key importance in qubit applications where the quantity of interest is the qubit state immediately at the start of the readout measurement. For example, Fig. \ref{fig:HMM} illustrates the predicted state-path of a 20 $\mu$s single shot using the forward-backward algorithm. The red line indicates the probability of being in the excited state $d_i=\ket{1}$ at each observation index, while the blue line represents the probability of being in the ground state $d_i=\ket{0}$ at each observation. It can been seen that for this single shot a relaxation transition is predicted near observation index $i =125$, yet the probability the qubit was in the excited state at the start of the readout measurement ($d_0 = \ket{1}$) is 99.98\%. 

\subsection{HMM Implementation}
The HMM readout scheme was implemented in Python with the Pomegranate package \cite{schreiber2017pomegranate}. Preparing a HMM in Pomegranate could be achieved by ``baking'' a model if the model parameters, $\lambda = ({\bf A},  B, \pi, N )$, were known. Alternatively, unsupervised training using the Baum Welch algorithm learned the model parameters from a dataset, but the number of states $N$ was required a-priori. Since the segments must be uncorrelated, preparing the heterodyned readout signal required careful selection of the segment demodulation time $\Delta t$. Hence, to find a suitable segment demodulation time we calculated the autocorrelation of the heterodyned readout signal and determined the point of minimum correlation. Although there were several points of minimum correlation, corresponding to the intermediate frequency ($\Omega_{IF} = 25$ MHz) of the signal, we sought the shortest possible time to ensure that the probability distributions formed by the discretized IQ observation segments, $B_d$, remained Gaussian. However, while training hidden Markov models, we found that the first minimum at 40 ns (one intermediate-frequency period) led to large statistical variations in the learned parameters. The next autocorrelation minimum which led to consistent HMM parameters corresponded to a segment demodulation time of $\Delta t = 80$ ns (two intermediate-frequency periods). 

The complete experimental dataset consisted of 25,250 single shots prepared in the ground state, and 25,250 shots prepared in the excited state. Each excited state measurement shot was taken with a pulse sequence consisting of a 25 ns $\pi$-pulse $R_x(\pi)$, followed by a $W = 20~ \mu$s rectangular readout pulse. The ground state shots were obtained with the same readout pulse but without any qubit excitation preceding it. The readout signal had a delay time of approximately 250 ns before the detection of the readout pulse, and another 250 ns was trimmed from the readout signal to ensure the readout cavity was in steady state. Therefore, the total delay between the qubit pulse and observations was 500 ns. Note that due to this delay we expect the assignment fidelity, defined by Eq. \eqref{eq:fidelity}, to be limited by the starting state population which is on the order of $\exp({-0.5/14.46}) = 96.6 \%$, where we used the effective relaxation time $T_{1,eff}=14.46 ~\mu$s as extracted from the HMM scheme (discussed below). After these adjustment, each single shot consisted of a sequence $\{O_i\}$ of 243 IQ observations (further details of our 3D cQED system can be found in \cite{wu2020high}).

\begin{figure}[t] 
\centering
\includegraphics[width=.4\textwidth]{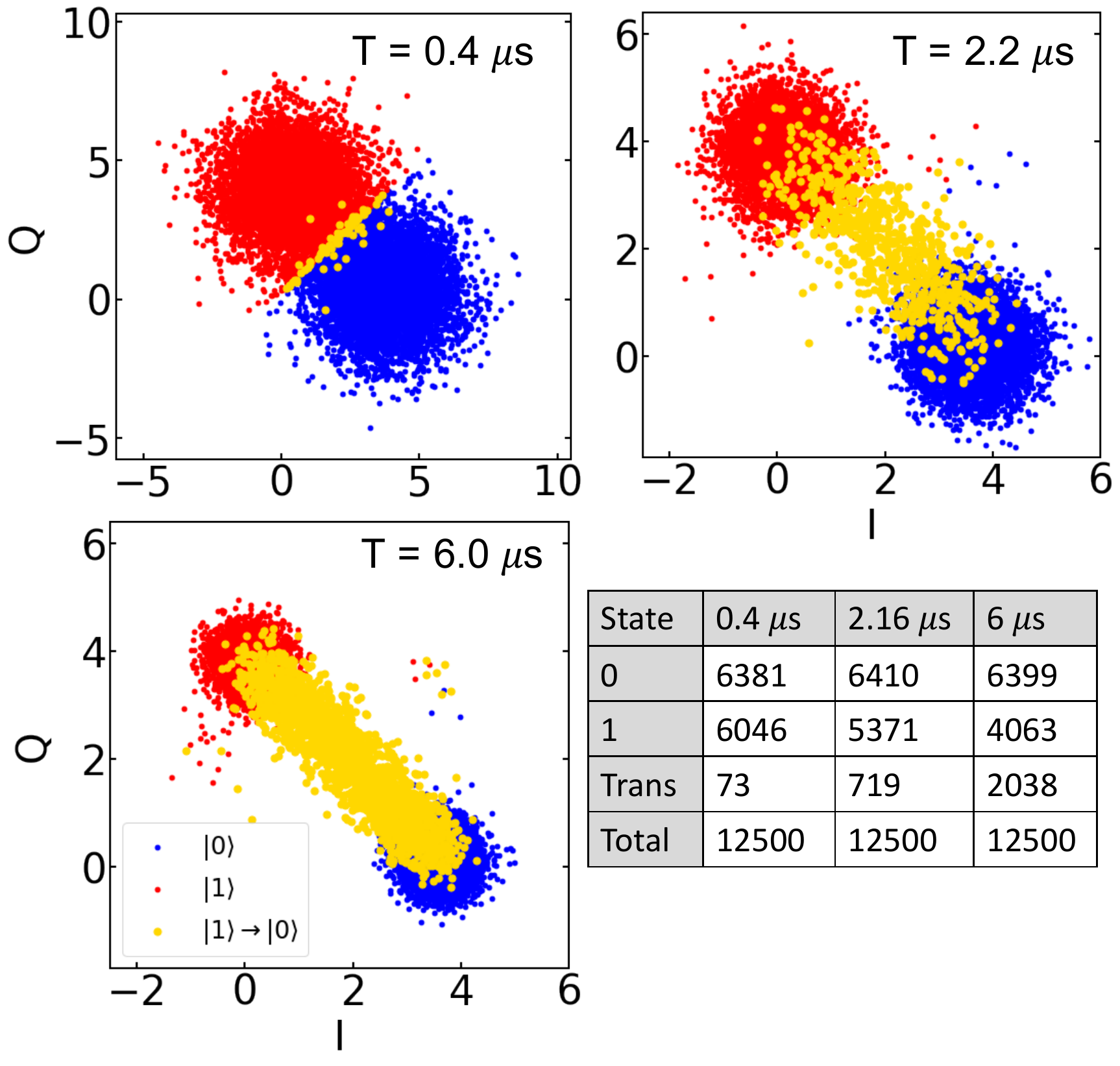}
\caption{Single shot state discrimination with a hidden Markov model (HMM) initial state classifier for three different readout times ($T$). A HMM is used to classify each shot into one of three cases; (blue) shots predicted to have started and remain in the ground state during the readout, (red) shots predicted to have started and remain in the excited state, and (yellow) shots in which a relaxation transition occurred during the readout. The table lists the counts.}
\label{fig:HMM_test}
\end{figure}

\subsection{Unsupervised Learning with HMMs}
Before proceeding to the main results we discuss the validation procedures used to confirm the reliability and consistency of hidden Markov models via Pomegranate. First, since qubit readout with HMMs gives access to the transition rates during the measurement, it is possible to extract the effective relaxation rate $T_{1,eff}$ under the influence of the readout signal. However, due to measurement induced decoherence the relaxation rate is not necessarily equal to the conventionally measured $T_1$ \cite{picot2008role,boissonneault2009dispersive} (operating in the strong projective regime). Therefore, to determine the accuracy in estimating $T_{1,eff}$ with unsupervised learning of HMMs, we generated 31 simulated datasets with the relaxation rates varying linearly from 1 $\mu$s to 16 $\mu$s. The learned relaxation rates were then calculated from the transition matrix element, $T_{1,eff} = (-80/\ln(a_{11}))$ ns, where $a_{11}$ was extracted from unsupervised learning using the Baum Welch algorithm. The standard deviation of the differences between the actual and learned $T_{1,eff}$ values was $0.175 ~\mu$s, and indicated that we could estimate the effective $T_{1,eff}$ to within 1.25\% in this range (see appendix). For our experimental dataset the learned $T_{1,eff}$ value was $(14.460 \pm .175) ~\mu$s. In comparison, measuring $T_1$ using the standard $R_x(\pi)\rightarrow variable~darktime\rightarrow readout ~pulse$ method resulted in $T_1 = 21 \pm 1 ~\mu$s. The difference of approximately 32\% between the conventionally measured $T_1$ and the learned value was consistent with the qubit induced dephasing \cite{picot2008role,boissonneault2009dispersive} caused by the readout amplitude of $\sim 20$ photons \cite{wu2020high}, estimated via a AC-Stark shift calibration \cite{schuster2005ac}. Similarly, the $\ket{0} \rightarrow \ket{1}$ excitation rate estimated from the $a_{01}$ transition matrix element was $\Gamma_{01} = 9.6\pm0.2$Hz, which is reasonable given previous residual excited-state population measurements \cite{jin2015thermal}. 

Next, bootstrap sampling techniques were used to extract the statistical variations in unsupervised training of HMMs via Pomegranate. The bootstrap technique consisted of generating 100 randomized subsets for each state from the experimental dataset. Each randomized bootstrapped subset consisted of a total of 2,000 ground state and 2,000 excited state single shot measurements. HMMs were then trained from each bootstrap subset using the Baum Welch algorithm. The standard deviation from the bootstrap technique on the learned transition matrix parameters was under $0.03\%$, and under $1\%$ for the means of the IQ distributions, indicating good consistency in unsupervised training of hidden Markov models via Pomegranate (see appendix for details).

\section{Main Results} 
\subsection{Hidden Markov Model State Classifier}
The HMM classifier was implemented by using unsupervised learning on a training data subset consisting of 2000 shots prepared in the ground state and 2000 shots prepared in the excited state. For single shot classification the starting state probabilities of the HMM were then modified such that $\pi_{\ket{0}}=\pi_{\ket{1}} = 0.50$. The classification scheme was based on the state-path predicted by the forward-backward algorithm, and a ``$0$'' or ``$1$'' label was assigned based on the state that had the maximum starting state probability. The remaining 46,500 shots were used as a test dataset. In Fig. \ref{fig:HMM_test}, 6,250 excited state shots and 6,250 ground state shots were classified with the HMM readout scheme for three different readout times. Shots that were predicted to start and remain in the excited state during the readout measurement were labeled in red, and those predicted to have started and remain in the ground state were labeled in blue. Most importantly, with the HMM readout scheme transitions can be detected with high certainty while maintaining high fidelity in the determination of the starting state of the measurement. This is illustrated by labeling in yellow those shots in which a transition was predicted. 

\begin{figure}[h]
   \centering
   \includegraphics[width = .35\textwidth]{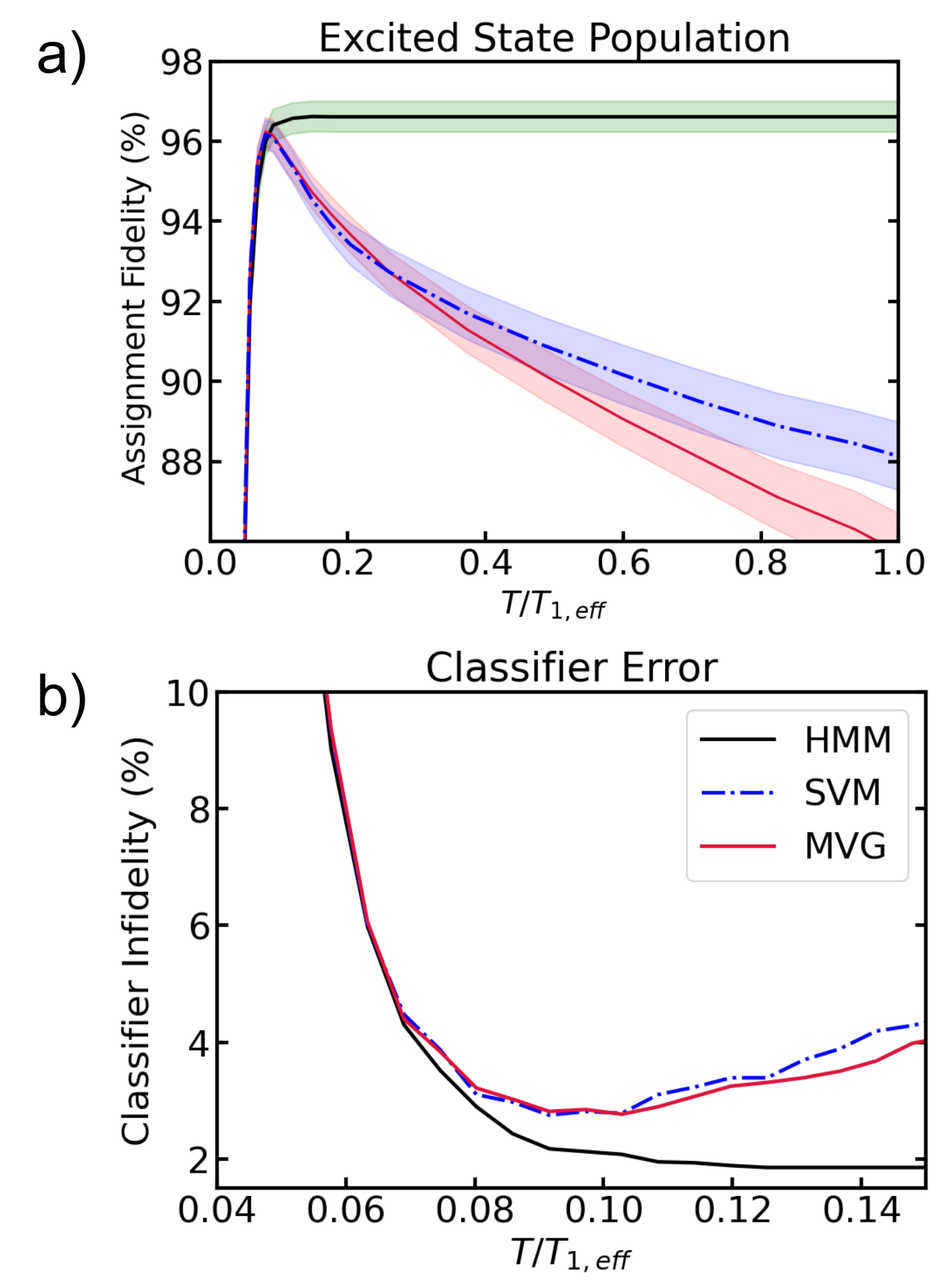}   
   \caption{a) Classifier fidelity in units of the effective relaxation time $T_{1,eff}$ for the excited state. The HMM scheme achieves a maximum assignment fidelity of $96.5\% \pm 0.4\%$, while the SVM and MVG methods achieve a fidelity of $95.9\%\pm0.4\%$, and $96.1\%\pm0.4\%$, respectively. Shaded regions indicate errors. The HMM scheme is robust against qubit relaxation time, eliminating the need for readout time optimization. b) Total classification error determined with a simulated dataset in which the starting state is prepared with 100\% preparation fidelity demonstrates that the HMM scheme has a lower total classification error.}
   \label{fig:clf_fidelities}
\end{figure}

Next, the excited state assignment fidelity ${\cal F}_{a,1} = 1-P(0|1)$ defined by equation \eqref{eq:fidelity} is compared for the HMM readout classifier against a multivariate Gaussian (MVG) classifier, and a support vector machines (SVM) classifier. Here the full dataset of 46,500 shots was classified, and errors for the HMM classifier were extracted from the bootstrap samples. Fig. \ref{fig:clf_fidelities}(a) shows the classification readout fidelity, ${\cal F}_{a,1}$, for the excited state as a function of the readout time in units of $T_{1,eff}$. The assignment fidelities for the HMM, SVM, and MVG classifiers were $96.5\% \pm 0.4\%$, $95.9\%\pm0.4\%$, and $96.1\%\pm0.4\%$, respectively. A striking difference between the datasets is that the HMM method is impervious to qubit state transitions. Since the HMM scheme can determine the starting state of shots that underwent a state transition with high probability, the readout fidelity remains fixed as a function of the readout time beyond $\sim 1~\mu$s. Note that for the HMM scheme the dominant source of misclassification was observed from shots that had a transitions within the first few observation segments. Since approximately 500 ns were trimmed from the start of each shot and state preparation errors were estimated to be less than 1\% \cite{wu2020high}, the excited state assignment fidelity is expected to be limited by the starting state population which is on the order of $\exp({-0.5/14.46}) = 96.6 \%$. This is consistent with the results presented in Fig. \ref{fig:clf_fidelities}(a).  

The single shot classification errors for the starting state were then extracted from a simulated dataset having 100\% preparation fidelity. The simulated dataset for the excited state was created by first generating state sequences (i.e. a sequence of ones and zeroes) having an exponential probability distribution with $T_{1,eff} = 14.46~\mu$s. Independent, identically distributed random samples of IQ values were then drawn from the learned multi-variate Gaussian distributions according to the randomly generated state sequences. The ground state shots were simulated, with no transitions, by random sampling from a multi-variate Gaussian distribution. A plot of the total classifier error extracted from the simulated dataset, $1- {\cal F}_a = (P(0|1) + P(1|0))/2$, is shown in Fig. \ref{fig:clf_fidelities}(b). This measure quantifies the errors in the classification of the single shot measurements for both the excited and ground states. It can be seen that overall the HMM classifier has a misclassification error under 2\% with a plateau of 1.86\%, whereas the MVG (SVM) method reaches a minimum error of 2.75\% (2.77\%) before increasing as a function of the readout time. 

\subsection{Ideal Fidelity and Single Shot Efficiency}
Gaussian distributions from the classified shots can be obtained by using state path prediction with the HMM scheme, a feature that is not possible with SVM or MVG based classifiers. In turn, this feature makes it is possible to compute the fidelity as defined from the integration of the overlap regions. This method of computing fidelity from the overlap differs from the assignment fidelity in that the former will yield the maximum fidelity achievable, whereas the latter will be limited by the starting state population. We may extract the post-filtered Gaussian IQ probability distributions as follows. Using state-path prediction with the forward-backward algorithm, each shot is demodulated until a transition is detected. When a transition is detected, the shot is split at the transition and demodulated in two sections. One section corresponds to when the qubit was predicted in the excited state for that shot, and the other corresponds to when the qubit was predicted in the ground state. In shots with no predicted transitions the signal is demodulated for the complete integration time $T_{int}$. Demodulating in this fashion eliminates averaging over transitions, and relies on the ability of the HMM scheme to correctly predict the qubit state at each observation index. If the HMM scheme provides accurate state discrimination, we expect the sampled distributions of many shots to result in Gaussian distributions of equal variance for a given integration time. Note that in the case when $T_{int} >> T_1$ this no longer holds. 

\begin{figure}[t]
\centering
\includegraphics[width = .4\textwidth]{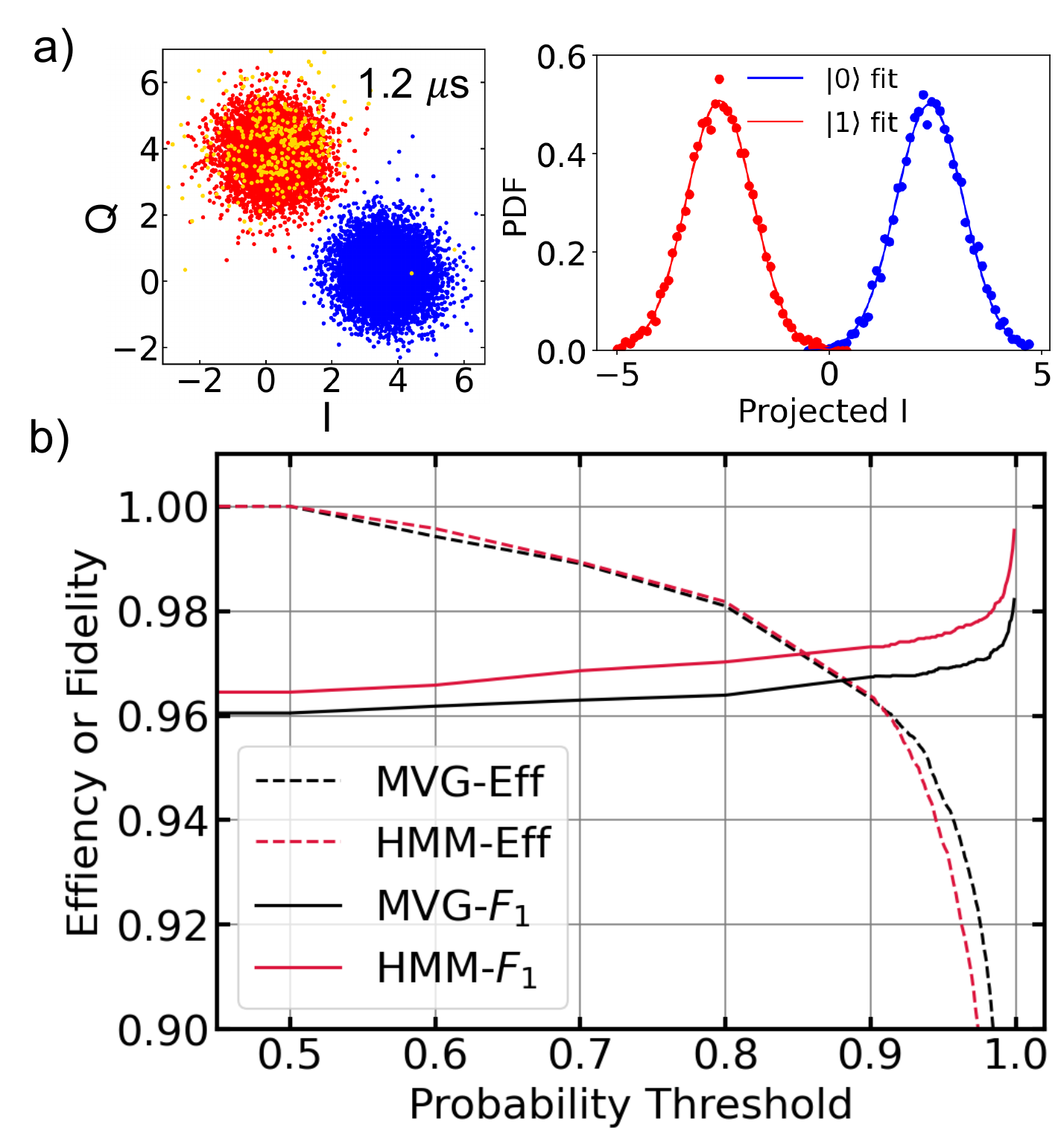}
\caption{a) IQ scatter plot and equal-variance Gaussian fits of ground (blue) and excited state (red) shots filtered with the HMM state discriminator for a 1.2 $\mu$s demodulation time. Shots predicted to start in the excited state which transitioned to the ground state during the readout are shown in yellow points on the left panel. Colored-matched markers on the right panel represent the projected distributions, solid lines represent the corresponding fits. Fitting errors are smaller than the markers. b) Fidelity may be improved by rejecting low probability shots via a threshold parameter, but only at the expense of readout efficiency.}
\label{fig:efficiency}
\end{figure}

The resulting IQ plot for $T_{int}=1.2~\mu$s is shown in Fig. \ref{fig:efficiency}(a). To compute the fidelity of the HMM filtered data, the HMM-filtered IQ distributions were projected onto the axis connecting the two centroids \cite{magesan2015machine}. Each of the projected distributions were then fitted simultaneously with equal-variance \textit{single} Gaussians. In contrast, the ideal fidelities were extracted by simultaneously fitting equal-variance \textit{double} Gaussians to each projected distribution of the unfiltered IQ data \cite{magesan2015machine}. Fidelities for both methods were then computed using equation \eqref{eq:ideal_fidelity}. Table I shows the results of the computed fidelities for various integration times ($T_{int}$), and shows that the HMM scheme reaches the ideal fidelity limit.

\begin{table}[h]
\begin{tabularx}{.49\textwidth}{llll}
\multicolumn{4}{c}{Table I. Maximum Fidelity from Gaussian Fits}\\
\hhline{====} \\
$T_{int}$ & $0.72\mu s$  & $1.2 \mu s$ & $2.16 \mu s$  \\
\hline \\
Ideal & ($99.14\pm.04$)\%  &($99.92\pm.02$)\%  & ($99.9987\pm .0005$)\%   \\
HMM &($99.12\pm .04$)\%  &($99.91\pm .06$)\%  &($99.998\pm .003$)\% \\
\hline
\end{tabularx}
\end{table}

The above method serves to illustrate the flexibility of using the HMM scheme by having access to state-path prediction for each shot. An alternative method to arrive at the same conclusion is by allowing low-probability shot rejection with the HMM classifier, which increases the readout fidelity at the expense of reducing the readout efficiency. The efficiency is quantified by the ratio of accepted shots versus attempted shots. Fig. \ref{fig:efficiency}(b) shows the efficiency and the excited state assignment fidelity for both, the HMM and MVG schemes. In this case we selected the optimal readout time of approximately $5\%\times T_{1_eff}$ for the MVG classifier and a arbitrary time of $10\%\times T_{1,eff}$ for the HMM scheme. It can be seen the HMM scheme achieved a higher assignment fidelity than the MVG scheme, while maintaining a comparable efficiency. We omit the SVM method since its classification scheme is based on a geometric approach and, thus, does not enable low probability shot rejection.

\section{Conclusion}
We demonstrated that hidden Markov models allow for a robust state-path readout scheme via transition detection, thus, allowing for starting state determination with high fidelity. The HMM scheme also demonstrated consistent classification performance even for readout times comparable with qubit $T_1$ times, where in contrast, current state of the art schemes are hindered by qubit state decay. Thus, using the HMM readout scheme eliminates the need for optimizing the integration time, a process which must be tailored to each superconducting qubit due to individual variations in their $T_1$ times. Meanwhile, the state-path trajectory HMM scheme is compatible with real time control systems, e.g. quantum orchestration platforms, which can lead to measurement speed up. Furthermore, unsupervised learning with the Baum Welch algorithm provides a tool for learning about transitions rates between quantum states and distribution parameters, thus, giving easy access to information not accessible with current state-of-the-art qubit readout schemes. Indeed, owing to the Markovian nature of qubit relaxation, hidden Markov models are a natural platform for qubit readout, and which, can handle multi-level qudit systems as well.

We acknowledge the helpful conversations with Kater Murch, Xian Wu, Jacob Schreiber, and Jason Bernstein. This work was performed under the auspices of the U.S. Department of Energy by Lawrence Livermore National Laboratory under Contract DE-AC52-07NA27344. The authors gratefully acknowledge support from the Department of Energy Office of Advanced Scientific Computing Research, Quantum Testbed Pathfinder Program under Award 2017-LLNL-SCW163, the National Nuclear Security Administration Advanced Simulation and Computing Beyond Moore's Law program (LLNL-ABS- 795437) and Lab Directed Research and Development award LDRD19-ERD-013. (LLNL-JRNL-810931)

\appendix{Appendix}

\subsection{Bootstrap results}
\begin{figure}[h]
   \centering
   \includegraphics[width =.45\textwidth]{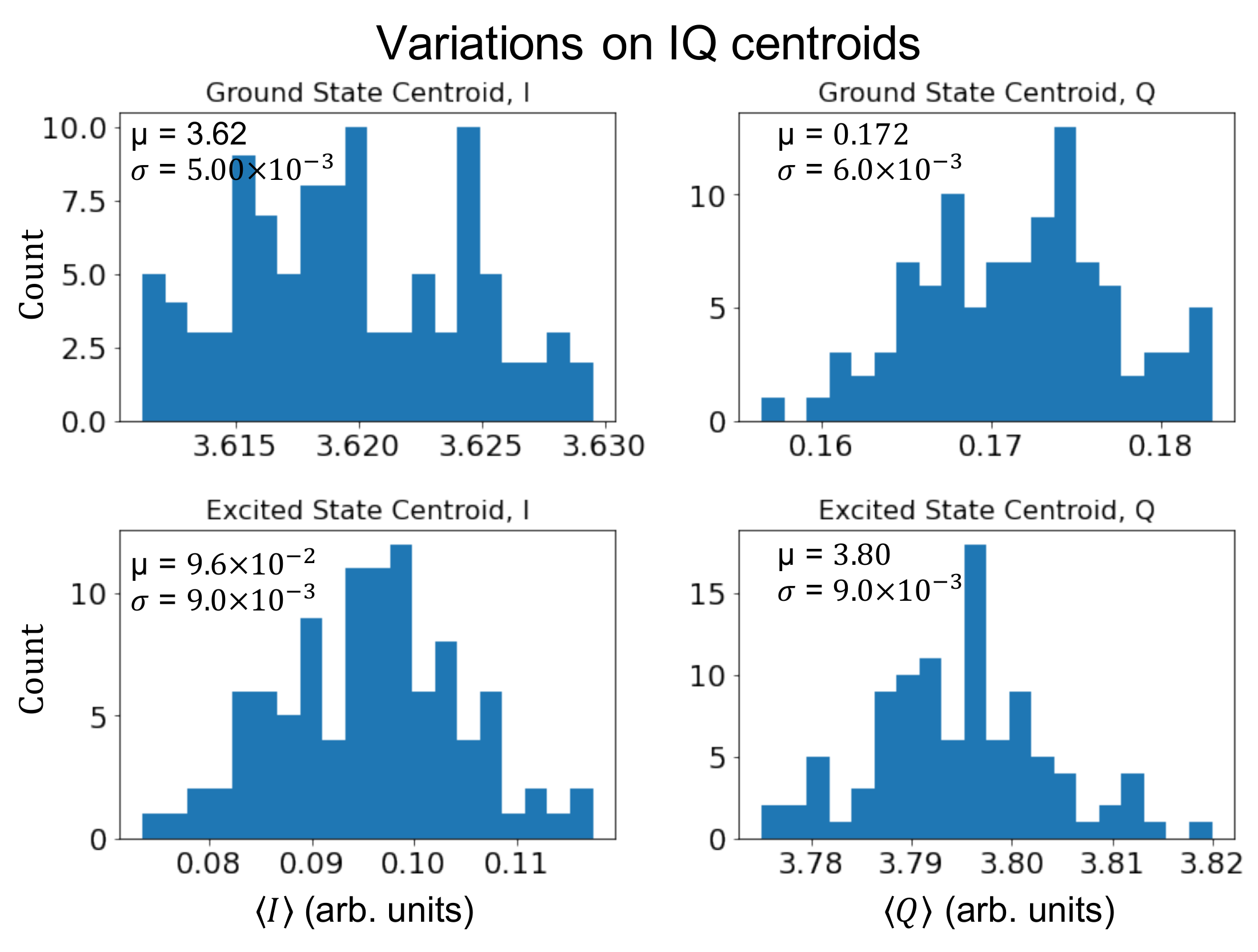} 
   \caption{Statistical variations extracted by the bootstrap method on the learned values of the means of the Gaussian IQ distributions for both the ground and excited state.}
   \label{fig:bootstrap_results_means}
\end{figure}

\begin{figure}[h]
   \centering
   \includegraphics[width =.45\textwidth]{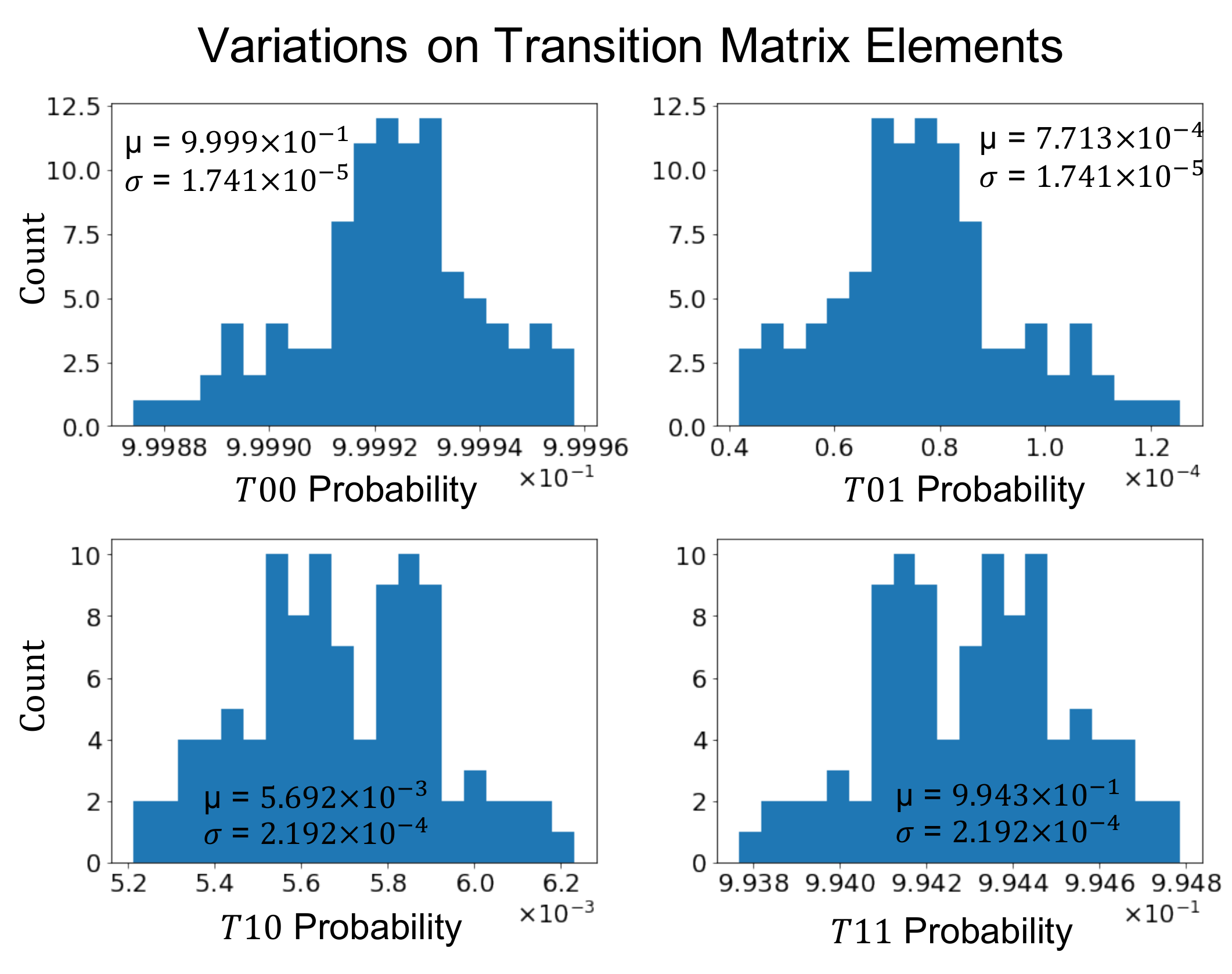} 
   \caption{Statistical variations extracted by the bootstrap technique on the learn values of the transition matrix (${\bf A}$).} 
   \label{fig:bootstrap_results_T}
\end{figure}

The statistical variations on unsupervised learning of the HMM parameters were obtained by using a bootstrap technique. For the bootstrap technique we randomly generated 100 bootstrap data subsets from the full experimental dataset. The sampling for each data subset was done with replacement, and each subset consisted of 2000 records in the ground state and 2000 records in the excited state. Each bootstrapped data subset was then used to train a hidden markov model (HMM) via the Baum Welch algorithm and statistics were generated on the variations on the learned parameters. The variations for the learned means of the IQ distributions are shown in Fig. \ref{fig:bootstrap_results_means} along with standard deviations. Similarly, the variations for the transition matrix elements are shown in Fig. \ref{fig:bootstrap_results_T}.

\subsection{HMM with 2D Gaussian distributions}\label{HMM_details}
This section is intended to illustrate the notation used in hidden Markov models (HMM) with an example. Suppose we have observed the following sequence of observations, $O_{0:T}=\{O_0,O_1,O_2,\dots,O_T\}$, and that there is a corresponding sequence of hidden states, $d_{0:T}=\{d_0,d_1,d_2,\dots,d_T\}$, from which these observations were emitted, probabilistically. Here, $d$ represents the hidden state explicitly, e.g. $d=\ket{0}$, or $d =\ket{1}$, and the index simply maintains the order in the sequence. In a HMM we wish to calculate the probability distributions over the many possible hidden-state configurations a sequence $O$ of length $T$ can be observed. This is achieved by computing the joint probability of a hidden-state sequence $d_{0:T}$ and its corresponding observation sequence $O_{0:T}$,
\begin{align}
&P(d_{0:T},O_{0:T}) = \nonumber \\
&P(d_0)P(O_0|d_0)P(d_1|d_0)P(d_2|d_1)P(O_1|d_1)P(O_2|d_2)\cdots\nonumber\\
&=  P(d_0)P(O_0|d_0)\displaystyle\prod_{i=1}^{i=T} P(d_i|d_{i-1})P(O_i|d_i).
\end{align}
In this notation $P(d_0)\equiv  \pi_{d_0}$ represents the probability of starting in state $d_0$. The conditional probability $P(d_i|d_{i-1})$ represents the probability of transitioning to state $d_i$ given we were in state $d_{i-1}$, thus, it defines the transition matrix ${\bf A}$. The conditional probability $P(O_i|d_i)$, known as the emission distribution in the formalism of HMMs, is the probability of observation $O_i$ given state $d_i$. For this example we assume the emission distributions are represented by 2-dimensional equal-variance ($\sigma^2$) Gaussians, therefore, the emission distributions ($B_d$) are given by
\begin{align}
&B_d(I,Q)\equiv P(O_i|d_i)\nonumber \\
& = \frac{1}{2\pi\sigma^2}\exp\left(-\frac{1}{2\sigma^2}[(I-\bar{I}_{d_i})^2 + (Q-\bar{Q}_{d_i})^2] \right),
\end{align}
where $\bar{I},\bar{Q}$ represent the center of the Gaussian distribution for state $d_i$, and the observation $O_i$ is defined by the $(I,Q)$ coordinate that results from a single shot measurement. For a strong-projective single-shot qubit readout measurement the IQ coordinate is the result of the demodulated heterodyne signal, where $I$ represents the in-phase component and $Q$ the out-of-phase component. In general the emission distributions need not be Gaussian, and can in principle be modeled by various probability distributions. 

\begin{figure}[t]
   \centering
   \includegraphics[width =.45\textwidth]{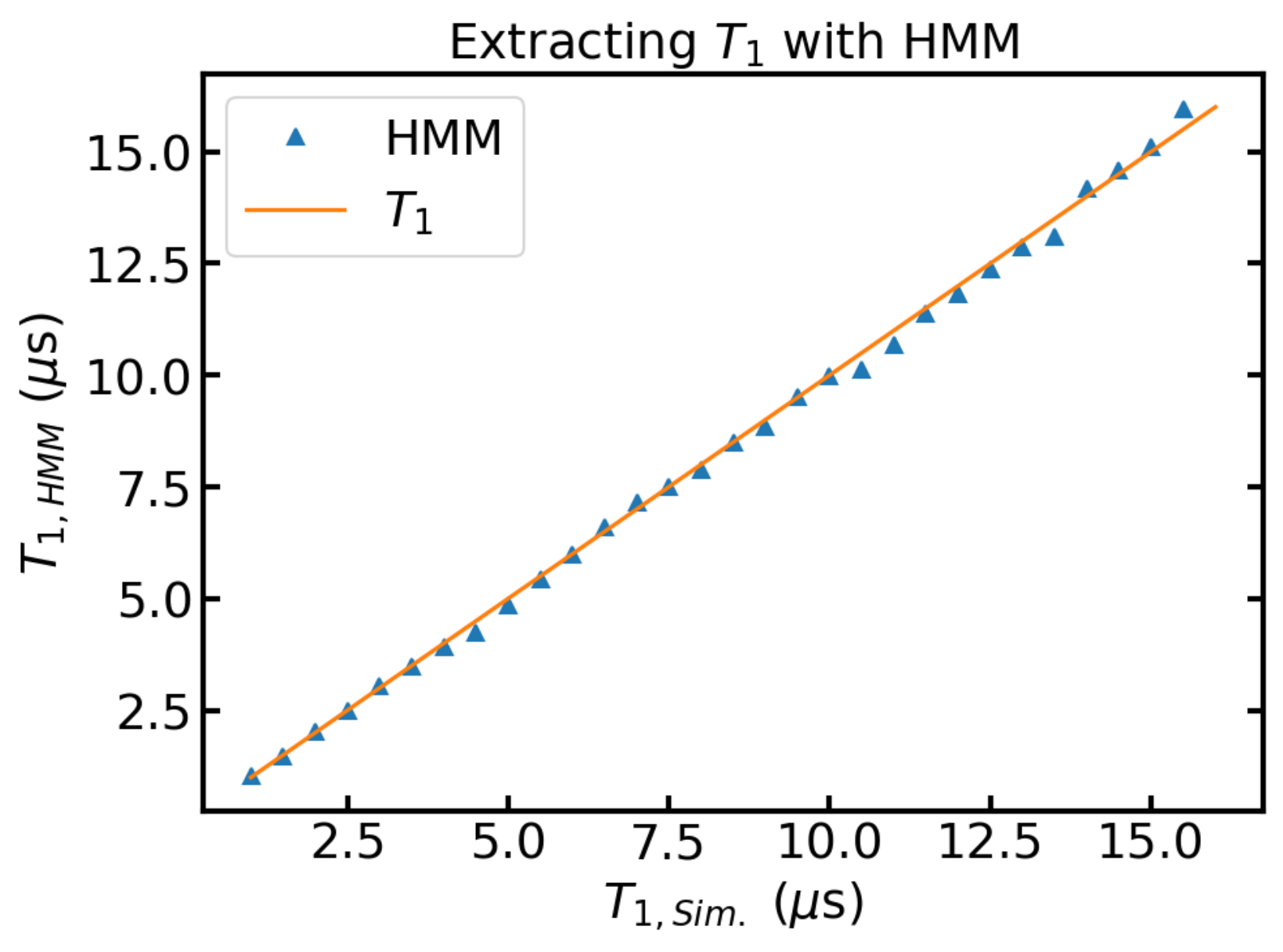} 
   \caption{Simulated datasets varying $T_1$ from 1 to 16 microseconds were used to trained several HMMs via unsupervised learning using the Baum Welch algorithm. The y-axis represents the $T_1$ values extracted from HMM and they are plotted against the actual simulated values.}
   \label{fig:T1_accuracy}
\end{figure}

\subsection{Verifying $T_1$ accuracy with HMM}
To extract the accuracy in estimating the effective $T_1$ value with HMM, we generated several simulated datasets with $T_1$ times varying linearly from $1\mu$s to $16\mu$s. Unsupervised learning using the Baum Welch algorithm was used to learn the parameters, and the $T_1$ was estimated from the transition matrix element $a_{11} = exp(-\Delta t/T_{1,HMM})$,
\begin{align}
T_{1,HMM} = -\frac{80}{\ln(a_{11})}~ns,
\end{align}
where we set $\Delta t = 80$ ns, which corresponds to $80~\text{ns} \times 2~\text{GHz} = 160$ samples for our 2 GHz ADC. The results are shown in Fig. \ref{fig:T1_accuracy}. The standard deviation of the differences (normalized) was $1.25\%$. The mean deviation between the actual and learned values was 0.0576 ns with a standard deviation of 0.175 ns. This indicates that we can estimate $T_1$ to within 1.25\%. Fitting a linear function $y=mx$ gives $m = 0.994 \pm .003$, which is a difference of $m_{exp} - m_{theo} =1 - 0.994= .006$, showing good stability over the entire sampled range. 

\subsection{Confirming $T_{1,eff}$ via starting state probabilities}
\begin{figure}[h]
   \centering
   \includegraphics[width=.45\textwidth]{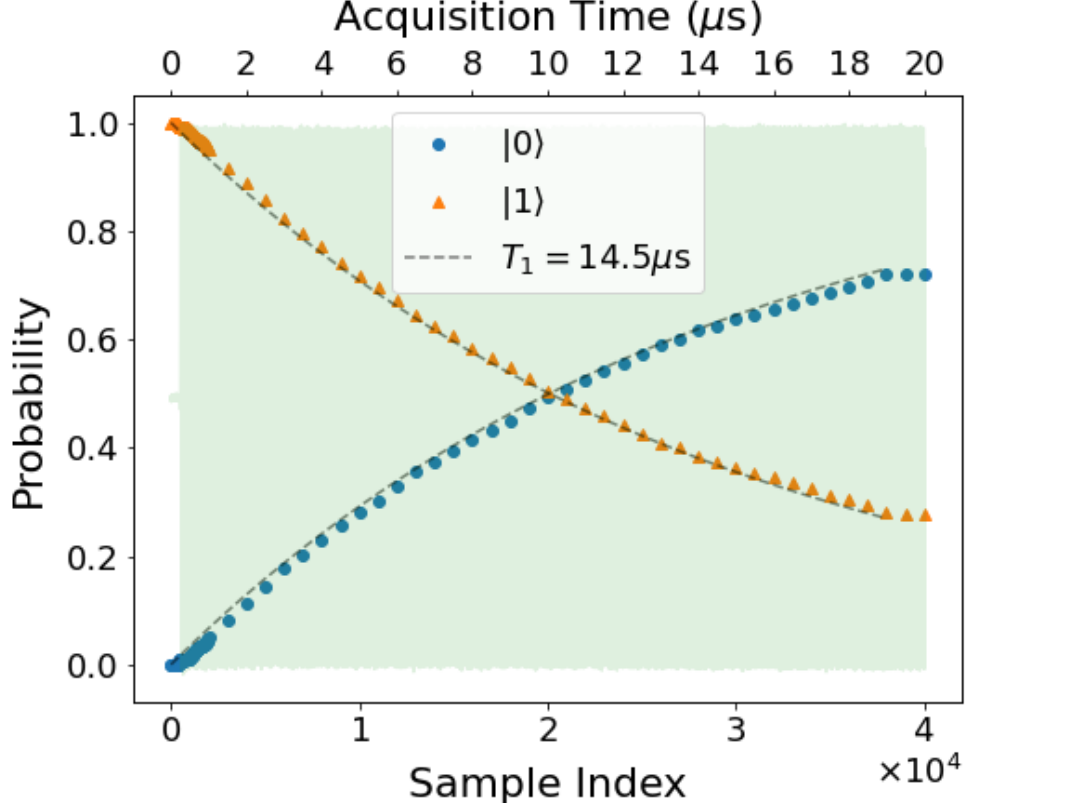} 
   \caption{The effective $T_1$ time extracted from the unsupervised learning of the starting state distributions by sweeping the start time of the integration during the demodulation.}
   \label{fig:T1eff}
\end{figure}

Due to measurement induced dephasing from the strong readout pulse, the $T_1$ extracted during a readout measurement is expected to be reduced from the typical measured $T_1$ value, i.e. not during the readout pulse. As an alternative to using the transition matrix to calculate the effective $T_1$, i.e. during the readout measurement, we confirmed the value by using unsupervised learning with HMMs to extract the starting state distributions $\pi = (\pi_{\ket{0}},\pi_{\ket{1}})$, i.e. the priors. This was accomplished by varying the demodulation start time (i.e. starting point of integration) for the experimental dataset while extracting the starting state distributions with HMM. This is time consuming, but uses the HMM's optimization to calculate the best fit for the model parameters $({\bf A},B,\pi,N)$ and learn the starting state distributions. Fig. \ref{fig:T1eff}shows the results along with the fitted exponential.
 
\begin{thebibliography}{34}%
\makeatletter
\providecommand \@ifxundefined [1]{%
 \@ifx{#1\undefined}
}%
\providecommand \@ifnum [1]{%
 \ifnum #1\expandafter \@firstoftwo
 \else \expandafter \@secondoftwo
 \fi
}%
\providecommand \@ifx [1]{%
 \ifx #1\expandafter \@firstoftwo
 \else \expandafter \@secondoftwo
 \fi
}%
\providecommand \natexlab [1]{#1}%
\providecommand \enquote  [1]{``#1''}%
\providecommand \bibnamefont  [1]{#1}%
\providecommand \bibfnamefont [1]{#1}%
\providecommand \citenamefont [1]{#1}%
\providecommand \href@noop [0]{\@secondoftwo}%
\providecommand \href [0]{\begingroup \@sanitize@url \@href}%
\providecommand \@href[1]{\@@startlink{#1}\@@href}%
\providecommand \@@href[1]{\endgroup#1\@@endlink}%
\providecommand \@sanitize@url [0]{\catcode `\\12\catcode `\$12\catcode
  `\&12\catcode `\#12\catcode `\^12\catcode `\_12\catcode `\%12\relax}%
\providecommand \@@startlink[1]{}%
\providecommand \@@endlink[0]{}%
\providecommand \url  [0]{\begingroup\@sanitize@url \@url }%
\providecommand \@url [1]{\endgroup\@href {#1}{\urlprefix }}%
\providecommand \urlprefix  [0]{URL }%
\providecommand \Eprint [0]{\href }%
\providecommand \doibase [0]{http://dx.doi.org/}%
\providecommand \selectlanguage [0]{\@gobble}%
\providecommand \bibinfo  [0]{\@secondoftwo}%
\providecommand \bibfield  [0]{\@secondoftwo}%
\providecommand \translation [1]{[#1]}%
\providecommand \BibitemOpen [0]{}%
\providecommand \bibitemStop [0]{}%
\providecommand \bibitemNoStop [0]{.\EOS\space}%
\providecommand \EOS [0]{\spacefactor3000\relax}%
\providecommand \BibitemShut  [1]{\csname bibitem#1\endcsname}%
\let\auto@bib@innerbib\@empty
\bibitem [{\citenamefont {Barends}\ \emph {et~al.}(2015)\citenamefont
  {Barends}, \citenamefont {Lamata}, \citenamefont {Kelly}, \citenamefont
  {Garc{\'\i}a-{\'A}lvarez}, \citenamefont {Fowler}, \citenamefont {Megrant},
  \citenamefont {Jeffrey}, \citenamefont {White}, \citenamefont {Sank},
  \citenamefont {Mutus} \emph {et~al.}}]{barends2015digital}%
  \BibitemOpen
  \bibfield  {author} {\bibinfo {author} {\bibfnamefont {R.}~\bibnamefont
  {Barends}}, \bibinfo {author} {\bibfnamefont {L.}~\bibnamefont {Lamata}},
  \bibinfo {author} {\bibfnamefont {J.}~\bibnamefont {Kelly}}, \bibinfo
  {author} {\bibfnamefont {L.}~\bibnamefont {Garc{\'\i}a-{\'A}lvarez}},
  \bibinfo {author} {\bibfnamefont {A.}~\bibnamefont {Fowler}}, \bibinfo
  {author} {\bibfnamefont {A.}~\bibnamefont {Megrant}}, \bibinfo {author}
  {\bibfnamefont {E.}~\bibnamefont {Jeffrey}}, \bibinfo {author} {\bibfnamefont
  {T.}~\bibnamefont {White}}, \bibinfo {author} {\bibfnamefont
  {D.}~\bibnamefont {Sank}}, \bibinfo {author} {\bibfnamefont {J.}~\bibnamefont
  {Mutus}},  \emph {et~al.},\ }\href@noop {} {\bibfield  {journal} {\bibinfo
  {journal} {Nature communications}\ }\textbf {\bibinfo {volume} {6}},\
  \bibinfo {pages} {1} (\bibinfo {year} {2015})}\BibitemShut {NoStop}%
\bibitem [{\citenamefont {Wendin}(2017)}]{wendin2017quantum}%
  \BibitemOpen
  \bibfield  {author} {\bibinfo {author} {\bibfnamefont {G.}~\bibnamefont
  {Wendin}},\ }\href@noop {} {\bibfield  {journal} {\bibinfo  {journal}
  {Reports on Progress in Physics}\ }\textbf {\bibinfo {volume} {80}},\
  \bibinfo {pages} {106001} (\bibinfo {year} {2017})}\BibitemShut {NoStop}%
\bibitem [{\citenamefont {Colless}\ \emph {et~al.}(2018)\citenamefont
  {Colless}, \citenamefont {Ramasesh}, \citenamefont {Dahlen}, \citenamefont
  {Blok}, \citenamefont {Kimchi-Schwartz}, \citenamefont {McClean},
  \citenamefont {Carter}, \citenamefont {de~Jong},\ and\ \citenamefont
  {Siddiqi}}]{colless2018computation}%
  \BibitemOpen
  \bibfield  {author} {\bibinfo {author} {\bibfnamefont {J.~I.}\ \bibnamefont
  {Colless}}, \bibinfo {author} {\bibfnamefont {V.~V.}\ \bibnamefont
  {Ramasesh}}, \bibinfo {author} {\bibfnamefont {D.}~\bibnamefont {Dahlen}},
  \bibinfo {author} {\bibfnamefont {M.~S.}\ \bibnamefont {Blok}}, \bibinfo
  {author} {\bibfnamefont {M.~E.}\ \bibnamefont {Kimchi-Schwartz}}, \bibinfo
  {author} {\bibfnamefont {J.~R.}\ \bibnamefont {McClean}}, \bibinfo {author}
  {\bibfnamefont {J.}~\bibnamefont {Carter}}, \bibinfo {author} {\bibfnamefont
  {W.~A.}\ \bibnamefont {de~Jong}}, \ and\ \bibinfo {author} {\bibfnamefont
  {I.}~\bibnamefont {Siddiqi}},\ }\href@noop {} {\bibfield  {journal} {\bibinfo
   {journal} {Physical Review X}\ }\textbf {\bibinfo {volume} {8}},\ \bibinfo
  {pages} {011021} (\bibinfo {year} {2018})}\BibitemShut {NoStop}%
\bibitem [{\citenamefont {Yan}\ \emph {et~al.}(2019)\citenamefont {Yan},
  \citenamefont {Zhang}, \citenamefont {Gong}, \citenamefont {Wu},
  \citenamefont {Zheng}, \citenamefont {Li}, \citenamefont {Wang},
  \citenamefont {Liang}, \citenamefont {Lin}, \citenamefont {Xu} \emph
  {et~al.}}]{yan2019strongly}%
  \BibitemOpen
  \bibfield  {author} {\bibinfo {author} {\bibfnamefont {Z.}~\bibnamefont
  {Yan}}, \bibinfo {author} {\bibfnamefont {Y.-R.}\ \bibnamefont {Zhang}},
  \bibinfo {author} {\bibfnamefont {M.}~\bibnamefont {Gong}}, \bibinfo {author}
  {\bibfnamefont {Y.}~\bibnamefont {Wu}}, \bibinfo {author} {\bibfnamefont
  {Y.}~\bibnamefont {Zheng}}, \bibinfo {author} {\bibfnamefont
  {S.}~\bibnamefont {Li}}, \bibinfo {author} {\bibfnamefont {C.}~\bibnamefont
  {Wang}}, \bibinfo {author} {\bibfnamefont {F.}~\bibnamefont {Liang}},
  \bibinfo {author} {\bibfnamefont {J.}~\bibnamefont {Lin}}, \bibinfo {author}
  {\bibfnamefont {Y.}~\bibnamefont {Xu}},  \emph {et~al.},\ }\href@noop {}
  {\bibfield  {journal} {\bibinfo  {journal} {Science}\ }\textbf {\bibinfo
  {volume} {364}},\ \bibinfo {pages} {753} (\bibinfo {year}
  {2019})}\BibitemShut {NoStop}%
\bibitem [{\citenamefont {Arute}\ \emph {et~al.}(2019)\citenamefont {Arute},
  \citenamefont {Arya}, \citenamefont {Babbush}, \citenamefont {Bacon},
  \citenamefont {Bardin}, \citenamefont {Barends}, \citenamefont {Biswas},
  \citenamefont {Boixo}, \citenamefont {Brandao}, \citenamefont {Buell} \emph
  {et~al.}}]{arute2019quantum}%
  \BibitemOpen
  \bibfield  {author} {\bibinfo {author} {\bibfnamefont {F.}~\bibnamefont
  {Arute}}, \bibinfo {author} {\bibfnamefont {K.}~\bibnamefont {Arya}},
  \bibinfo {author} {\bibfnamefont {R.}~\bibnamefont {Babbush}}, \bibinfo
  {author} {\bibfnamefont {D.}~\bibnamefont {Bacon}}, \bibinfo {author}
  {\bibfnamefont {J.~C.}\ \bibnamefont {Bardin}}, \bibinfo {author}
  {\bibfnamefont {R.}~\bibnamefont {Barends}}, \bibinfo {author} {\bibfnamefont
  {R.}~\bibnamefont {Biswas}}, \bibinfo {author} {\bibfnamefont
  {S.}~\bibnamefont {Boixo}}, \bibinfo {author} {\bibfnamefont {F.~G.}\
  \bibnamefont {Brandao}}, \bibinfo {author} {\bibfnamefont {D.~A.}\
  \bibnamefont {Buell}},  \emph {et~al.},\ }\href@noop {} {\bibfield  {journal}
  {\bibinfo  {journal} {Nature}\ }\textbf {\bibinfo {volume} {574}},\ \bibinfo
  {pages} {505} (\bibinfo {year} {2019})}\BibitemShut {NoStop}%
\bibitem [{\citenamefont {Lucero}\ \emph {et~al.}(2008)\citenamefont {Lucero},
  \citenamefont {Hofheinz}, \citenamefont {Ansmann}, \citenamefont {Bialczak},
  \citenamefont {Katz}, \citenamefont {Neeley}, \citenamefont {OConnell},
  \citenamefont {Wang}, \citenamefont {Cleland},\ and\ \citenamefont
  {Martinis}}]{lucero2008high}%
  \BibitemOpen
  \bibfield  {author} {\bibinfo {author} {\bibfnamefont {E.}~\bibnamefont
  {Lucero}}, \bibinfo {author} {\bibfnamefont {M.}~\bibnamefont {Hofheinz}},
  \bibinfo {author} {\bibfnamefont {M.}~\bibnamefont {Ansmann}}, \bibinfo
  {author} {\bibfnamefont {R.~C.}\ \bibnamefont {Bialczak}}, \bibinfo {author}
  {\bibfnamefont {N.}~\bibnamefont {Katz}}, \bibinfo {author} {\bibfnamefont
  {M.}~\bibnamefont {Neeley}}, \bibinfo {author} {\bibfnamefont {A.~D.}\
  \bibnamefont {OConnell}}, \bibinfo {author} {\bibfnamefont {H.}~\bibnamefont
  {Wang}}, \bibinfo {author} {\bibfnamefont {A.~N.}\ \bibnamefont {Cleland}}, \
  and\ \bibinfo {author} {\bibfnamefont {J.~M.}\ \bibnamefont {Martinis}},\
  }\href@noop {} {\bibfield  {journal} {\bibinfo  {journal} {Physical review
  letters}\ }\textbf {\bibinfo {volume} {100}},\ \bibinfo {pages} {247001}
  (\bibinfo {year} {2008})}\BibitemShut {NoStop}%
\bibitem [{\citenamefont {Rol}\ \emph {et~al.}(2017)\citenamefont {Rol},
  \citenamefont {Bultink}, \citenamefont {OBrien}, \citenamefont {de~Jong},
  \citenamefont {Theis}, \citenamefont {Fu}, \citenamefont {Luthi},
  \citenamefont {Vermeulen}, \citenamefont {de~Sterke}, \citenamefont {Bruno},
  \citenamefont {Deurloo}, \citenamefont {Schouten}, \citenamefont {Wilhelm},\
  and\ \citenamefont {DiCarlo}}]{rol2017restless}%
  \BibitemOpen
  \bibfield  {author} {\bibinfo {author} {\bibfnamefont {M.~A.}\ \bibnamefont
  {Rol}}, \bibinfo {author} {\bibfnamefont {C.~C.}\ \bibnamefont {Bultink}},
  \bibinfo {author} {\bibfnamefont {T.~E.}\ \bibnamefont {OBrien}}, \bibinfo
  {author} {\bibfnamefont {S.~R.}\ \bibnamefont {de~Jong}}, \bibinfo {author}
  {\bibfnamefont {L.~S.}\ \bibnamefont {Theis}}, \bibinfo {author}
  {\bibfnamefont {X.}~\bibnamefont {Fu}}, \bibinfo {author} {\bibfnamefont
  {F.}~\bibnamefont {Luthi}}, \bibinfo {author} {\bibfnamefont {R.~F.~L.}\
  \bibnamefont {Vermeulen}}, \bibinfo {author} {\bibfnamefont {J.~C.}\
  \bibnamefont {de~Sterke}}, \bibinfo {author} {\bibfnamefont {A.}~\bibnamefont
  {Bruno}}, \bibinfo {author} {\bibfnamefont {D.}~\bibnamefont {Deurloo}},
  \bibinfo {author} {\bibfnamefont {R.~N.}\ \bibnamefont {Schouten}}, \bibinfo
  {author} {\bibfnamefont {F.~K.}\ \bibnamefont {Wilhelm}}, \ and\ \bibinfo
  {author} {\bibfnamefont {L.}~\bibnamefont {DiCarlo}},\ }\href@noop {}
  {\bibfield  {journal} {\bibinfo  {journal} {Physical Review Applied}\
  }\textbf {\bibinfo {volume} {7}},\ \bibinfo {pages} {041001} (\bibinfo {year}
  {2017})}\BibitemShut {NoStop}%
\bibitem [{\citenamefont {Reed}\ \emph {et~al.}(2012)\citenamefont {Reed},
  \citenamefont {DiCarlo}, \citenamefont {Nigg}, \citenamefont {Sun},
  \citenamefont {Frunzio}, \citenamefont {Girvin},\ and\ \citenamefont
  {Schoelkopf}}]{reed2012realization}%
  \BibitemOpen
  \bibfield  {author} {\bibinfo {author} {\bibfnamefont {M.~D.}\ \bibnamefont
  {Reed}}, \bibinfo {author} {\bibfnamefont {L.}~\bibnamefont {DiCarlo}},
  \bibinfo {author} {\bibfnamefont {S.~E.}\ \bibnamefont {Nigg}}, \bibinfo
  {author} {\bibfnamefont {L.}~\bibnamefont {Sun}}, \bibinfo {author}
  {\bibfnamefont {L.}~\bibnamefont {Frunzio}}, \bibinfo {author} {\bibfnamefont
  {S.~M.}\ \bibnamefont {Girvin}}, \ and\ \bibinfo {author} {\bibfnamefont
  {R.~J.}\ \bibnamefont {Schoelkopf}},\ }\href@noop {} {\bibfield  {journal}
  {\bibinfo  {journal} {Nature}\ }\textbf {\bibinfo {volume} {482}},\ \bibinfo
  {pages} {382} (\bibinfo {year} {2012})}\BibitemShut {NoStop}%
\bibitem [{\citenamefont {C{\'o}rcoles}\ \emph {et~al.}(2015)\citenamefont
  {C{\'o}rcoles}, \citenamefont {Magesan}, \citenamefont {Srinivasan},
  \citenamefont {Cross}, \citenamefont {Steffen}, \citenamefont {Gambetta},\
  and\ \citenamefont {Chow}}]{corcoles2015demonstration}%
  \BibitemOpen
  \bibfield  {author} {\bibinfo {author} {\bibfnamefont {A.~D.}\ \bibnamefont
  {C{\'o}rcoles}}, \bibinfo {author} {\bibfnamefont {E.}~\bibnamefont
  {Magesan}}, \bibinfo {author} {\bibfnamefont {S.~J.}\ \bibnamefont
  {Srinivasan}}, \bibinfo {author} {\bibfnamefont {A.~W.}\ \bibnamefont
  {Cross}}, \bibinfo {author} {\bibfnamefont {M.}~\bibnamefont {Steffen}},
  \bibinfo {author} {\bibfnamefont {J.~M.}\ \bibnamefont {Gambetta}}, \ and\
  \bibinfo {author} {\bibfnamefont {J.~M.}\ \bibnamefont {Chow}},\ }\href@noop
  {} {\bibfield  {journal} {\bibinfo  {journal} {Nature communications}\
  }\textbf {\bibinfo {volume} {6}},\ \bibinfo {pages} {1} (\bibinfo {year}
  {2015})}\BibitemShut {NoStop}%
\bibitem [{\citenamefont {Walter}\ \emph {et~al.}(2017)\citenamefont {Walter},
  \citenamefont {Kurpiers}, \citenamefont {Gasparinetti}, \citenamefont
  {Magnard}, \citenamefont {Poto{\v{c}}nik}, \citenamefont {Salath{\'e}},
  \citenamefont {Pechal}, \citenamefont {Mondal}, \citenamefont {Oppliger},
  \citenamefont {Eichler} \emph {et~al.}}]{walter2017rapid}%
  \BibitemOpen
  \bibfield  {author} {\bibinfo {author} {\bibfnamefont {T.}~\bibnamefont
  {Walter}}, \bibinfo {author} {\bibfnamefont {P.}~\bibnamefont {Kurpiers}},
  \bibinfo {author} {\bibfnamefont {S.}~\bibnamefont {Gasparinetti}}, \bibinfo
  {author} {\bibfnamefont {P.}~\bibnamefont {Magnard}}, \bibinfo {author}
  {\bibfnamefont {A.}~\bibnamefont {Poto{\v{c}}nik}}, \bibinfo {author}
  {\bibfnamefont {Y.}~\bibnamefont {Salath{\'e}}}, \bibinfo {author}
  {\bibfnamefont {M.}~\bibnamefont {Pechal}}, \bibinfo {author} {\bibfnamefont
  {M.}~\bibnamefont {Mondal}}, \bibinfo {author} {\bibfnamefont
  {M.}~\bibnamefont {Oppliger}}, \bibinfo {author} {\bibfnamefont
  {C.}~\bibnamefont {Eichler}},  \emph {et~al.},\ }\href@noop {} {\bibfield
  {journal} {\bibinfo  {journal} {Physical Review Applied}\ }\textbf {\bibinfo
  {volume} {7}},\ \bibinfo {pages} {054020} (\bibinfo {year}
  {2017})}\BibitemShut {NoStop}%
\bibitem [{\citenamefont {Nersisyan}\ \emph {et~al.}(2019)\citenamefont
  {Nersisyan}, \citenamefont {Poletto}, \citenamefont {Alidoust}, \citenamefont
  {Manenti}, \citenamefont {Renzas}, \citenamefont {Bui}, \citenamefont {Vu},
  \citenamefont {Whyland}, \citenamefont {Mohan}, \citenamefont {Sete} \emph
  {et~al.}}]{nersisyan2019manufacturing}%
  \BibitemOpen
  \bibfield  {author} {\bibinfo {author} {\bibfnamefont {A.}~\bibnamefont
  {Nersisyan}}, \bibinfo {author} {\bibfnamefont {S.}~\bibnamefont {Poletto}},
  \bibinfo {author} {\bibfnamefont {N.}~\bibnamefont {Alidoust}}, \bibinfo
  {author} {\bibfnamefont {R.}~\bibnamefont {Manenti}}, \bibinfo {author}
  {\bibfnamefont {R.}~\bibnamefont {Renzas}}, \bibinfo {author} {\bibfnamefont
  {C.-V.}\ \bibnamefont {Bui}}, \bibinfo {author} {\bibfnamefont
  {K.}~\bibnamefont {Vu}}, \bibinfo {author} {\bibfnamefont {T.}~\bibnamefont
  {Whyland}}, \bibinfo {author} {\bibfnamefont {Y.}~\bibnamefont {Mohan}},
  \bibinfo {author} {\bibfnamefont {E.~A.}\ \bibnamefont {Sete}},  \emph
  {et~al.},\ }\href@noop {} {\bibfield  {journal} {\bibinfo  {journal} {arXiv
  preprint arXiv:1901.08042}\ } (\bibinfo {year} {2019})}\BibitemShut {NoStop}%
\bibitem [{\citenamefont {Place}\ \emph {et~al.}(2020)\citenamefont {Place},
  \citenamefont {Rodgers}, \citenamefont {Mundada}, \citenamefont {Smitham},
  \citenamefont {Fitzpatrick}, \citenamefont {Leng}, \citenamefont {Premkumar},
  \citenamefont {Bryon}, \citenamefont {Sussman}, \citenamefont {Cheng} \emph
  {et~al.}}]{place2020new}%
  \BibitemOpen
  \bibfield  {author} {\bibinfo {author} {\bibfnamefont {A.~P.}\ \bibnamefont
  {Place}}, \bibinfo {author} {\bibfnamefont {L.~V.}\ \bibnamefont {Rodgers}},
  \bibinfo {author} {\bibfnamefont {P.}~\bibnamefont {Mundada}}, \bibinfo
  {author} {\bibfnamefont {B.~M.}\ \bibnamefont {Smitham}}, \bibinfo {author}
  {\bibfnamefont {M.}~\bibnamefont {Fitzpatrick}}, \bibinfo {author}
  {\bibfnamefont {Z.}~\bibnamefont {Leng}}, \bibinfo {author} {\bibfnamefont
  {A.}~\bibnamefont {Premkumar}}, \bibinfo {author} {\bibfnamefont
  {J.}~\bibnamefont {Bryon}}, \bibinfo {author} {\bibfnamefont
  {S.}~\bibnamefont {Sussman}}, \bibinfo {author} {\bibfnamefont
  {G.}~\bibnamefont {Cheng}},  \emph {et~al.},\ }\href@noop {} {\bibfield
  {journal} {\bibinfo  {journal} {arXiv preprint arXiv:2003.00024}\ } (\bibinfo
  {year} {2020})}\BibitemShut {NoStop}%
\bibitem [{\citenamefont {Dassonneville}\ \emph {et~al.}(2020)\citenamefont
  {Dassonneville}, \citenamefont {Ramos}, \citenamefont {Milchakov},
  \citenamefont {Planat}, \citenamefont {Dumur}, \citenamefont {Foroughi},
  \citenamefont {Puertas}, \citenamefont {Leger}, \citenamefont {Bharadwaj},
  \citenamefont {Delaforce} \emph {et~al.}}]{dassonneville2020fast}%
  \BibitemOpen
  \bibfield  {author} {\bibinfo {author} {\bibfnamefont {R.}~\bibnamefont
  {Dassonneville}}, \bibinfo {author} {\bibfnamefont {T.}~\bibnamefont
  {Ramos}}, \bibinfo {author} {\bibfnamefont {V.}~\bibnamefont {Milchakov}},
  \bibinfo {author} {\bibfnamefont {L.}~\bibnamefont {Planat}}, \bibinfo
  {author} {\bibfnamefont {{\'E}.}~\bibnamefont {Dumur}}, \bibinfo {author}
  {\bibfnamefont {F.}~\bibnamefont {Foroughi}}, \bibinfo {author}
  {\bibfnamefont {J.}~\bibnamefont {Puertas}}, \bibinfo {author} {\bibfnamefont
  {S.}~\bibnamefont {Leger}}, \bibinfo {author} {\bibfnamefont
  {K.}~\bibnamefont {Bharadwaj}}, \bibinfo {author} {\bibfnamefont
  {J.}~\bibnamefont {Delaforce}},  \emph {et~al.},\ }\href@noop {} {\bibfield
  {journal} {\bibinfo  {journal} {Physical Review X}\ }\textbf {\bibinfo
  {volume} {10}},\ \bibinfo {pages} {011045} (\bibinfo {year}
  {2020})}\BibitemShut {NoStop}%
\bibitem [{\citenamefont {Magesan}\ \emph {et~al.}(2015)\citenamefont
  {Magesan}, \citenamefont {Gambetta}, \citenamefont {C{\'o}rcoles},\ and\
  \citenamefont {Chow}}]{magesan2015machine}%
  \BibitemOpen
  \bibfield  {author} {\bibinfo {author} {\bibfnamefont {E.}~\bibnamefont
  {Magesan}}, \bibinfo {author} {\bibfnamefont {J.~M.}\ \bibnamefont
  {Gambetta}}, \bibinfo {author} {\bibfnamefont {A.~D.}\ \bibnamefont
  {C{\'o}rcoles}}, \ and\ \bibinfo {author} {\bibfnamefont {J.~M.}\
  \bibnamefont {Chow}},\ }\href@noop {} {\bibfield  {journal} {\bibinfo
  {journal} {Physical review letters}\ }\textbf {\bibinfo {volume} {114}},\
  \bibinfo {pages} {200501} (\bibinfo {year} {2015})}\BibitemShut {NoStop}%
\bibitem [{\citenamefont {Seif}\ \emph {et~al.}(2018)\citenamefont {Seif},
  \citenamefont {Landsman}, \citenamefont {Linke}, \citenamefont {Figgatt},
  \citenamefont {Monroe},\ and\ \citenamefont {Hafezi}}]{seif2018machine}%
  \BibitemOpen
  \bibfield  {author} {\bibinfo {author} {\bibfnamefont {A.}~\bibnamefont
  {Seif}}, \bibinfo {author} {\bibfnamefont {K.~A.}\ \bibnamefont {Landsman}},
  \bibinfo {author} {\bibfnamefont {N.~M.}\ \bibnamefont {Linke}}, \bibinfo
  {author} {\bibfnamefont {C.}~\bibnamefont {Figgatt}}, \bibinfo {author}
  {\bibfnamefont {C.}~\bibnamefont {Monroe}}, \ and\ \bibinfo {author}
  {\bibfnamefont {M.}~\bibnamefont {Hafezi}},\ }\href@noop {} {\bibfield
  {journal} {\bibinfo  {journal} {Journal of Physics B: Atomic, Molecular and
  Optical Physics}\ }\textbf {\bibinfo {volume} {51}},\ \bibinfo {pages}
  {174006} (\bibinfo {year} {2018})}\BibitemShut {NoStop}%
\bibitem [{\citenamefont {Rabiner}\ and\ \citenamefont
  {Juang}(1986)}]{rabiner1986introduction}%
  \BibitemOpen
  \bibfield  {author} {\bibinfo {author} {\bibfnamefont {L.}~\bibnamefont
  {Rabiner}}\ and\ \bibinfo {author} {\bibfnamefont {B.}~\bibnamefont
  {Juang}},\ }\href@noop {} {\bibfield  {journal} {\bibinfo  {journal} {ieee
  assp magazine}\ }\textbf {\bibinfo {volume} {3}},\ \bibinfo {pages} {4}
  (\bibinfo {year} {1986})}\BibitemShut {NoStop}%
\bibitem [{\citenamefont {Blais}\ \emph {et~al.}(2004)\citenamefont {Blais},
  \citenamefont {Huang}, \citenamefont {Wallraff}, \citenamefont {Girvin},\
  and\ \citenamefont {Schoelkopf}}]{blais2004cavity}%
  \BibitemOpen
  \bibfield  {author} {\bibinfo {author} {\bibfnamefont {A.}~\bibnamefont
  {Blais}}, \bibinfo {author} {\bibfnamefont {R.-S.}\ \bibnamefont {Huang}},
  \bibinfo {author} {\bibfnamefont {A.}~\bibnamefont {Wallraff}}, \bibinfo
  {author} {\bibfnamefont {S.~M.}\ \bibnamefont {Girvin}}, \ and\ \bibinfo
  {author} {\bibfnamefont {R.~J.}\ \bibnamefont {Schoelkopf}},\ }\href@noop {}
  {\bibfield  {journal} {\bibinfo  {journal} {Physical Review A}\ }\textbf
  {\bibinfo {volume} {69}},\ \bibinfo {pages} {062320} (\bibinfo {year}
  {2004})}\BibitemShut {NoStop}%
\bibitem [{\citenamefont {Cortes}\ and\ \citenamefont
  {Vapnik}(1995)}]{cortes1995support}%
  \BibitemOpen
  \bibfield  {author} {\bibinfo {author} {\bibfnamefont {C.}~\bibnamefont
  {Cortes}}\ and\ \bibinfo {author} {\bibfnamefont {V.}~\bibnamefont
  {Vapnik}},\ }\href@noop {} {\bibfield  {journal} {\bibinfo  {journal}
  {Machine learning}\ }\textbf {\bibinfo {volume} {20}},\ \bibinfo {pages}
  {273} (\bibinfo {year} {1995})}\BibitemShut {NoStop}%
\bibitem [{\citenamefont {Sun}\ and\ \citenamefont
  {Geller}(2018)}]{sun2018efficient}%
  \BibitemOpen
  \bibfield  {author} {\bibinfo {author} {\bibfnamefont {M.}~\bibnamefont
  {Sun}}\ and\ \bibinfo {author} {\bibfnamefont {M.~R.}\ \bibnamefont
  {Geller}},\ }\href@noop {} {\bibfield  {journal} {\bibinfo  {journal} {arXiv
  preprint arXiv:1810.10523}\ } (\bibinfo {year} {2018})}\BibitemShut {NoStop}%
\bibitem [{\citenamefont {Geller}(2020)}]{geller2020rigorous}%
  \BibitemOpen
  \bibfield  {author} {\bibinfo {author} {\bibfnamefont {M.~R.}\ \bibnamefont
  {Geller}},\ }\href@noop {} {\bibfield  {journal} {\bibinfo  {journal}
  {Quantum Science and Technology}\ }\textbf {\bibinfo {volume} {5}},\ \bibinfo
  {pages} {03LT01} (\bibinfo {year} {2020})}\BibitemShut {NoStop}%
\bibitem [{\citenamefont {Wu}\ \emph {et~al.}(2020)\citenamefont {Wu},
  \citenamefont {Tomarken}, \citenamefont {Petersson}, \citenamefont
  {Martinez}, \citenamefont {Rosen},\ and\ \citenamefont
  {DuBois}}]{wu2020high}%
  \BibitemOpen
  \bibfield  {author} {\bibinfo {author} {\bibfnamefont {X.}~\bibnamefont
  {Wu}}, \bibinfo {author} {\bibfnamefont {S.~L.}\ \bibnamefont {Tomarken}},
  \bibinfo {author} {\bibfnamefont {N.~A.}\ \bibnamefont {Petersson}}, \bibinfo
  {author} {\bibfnamefont {L.~A.}\ \bibnamefont {Martinez}}, \bibinfo {author}
  {\bibfnamefont {Y.~J.}\ \bibnamefont {Rosen}}, \ and\ \bibinfo {author}
  {\bibfnamefont {J.~L.}\ \bibnamefont {DuBois}},\ }\href {\doibase
  10.1103/PhysRevLett.125.170502} {\bibfield  {journal} {\bibinfo  {journal}
  {Phys. Rev. Lett.}\ }\textbf {\bibinfo {volume} {125}},\ \bibinfo {pages}
  {170502} (\bibinfo {year} {2020})}\BibitemShut {NoStop}%
\bibitem [{\citenamefont {Wallraff}\ \emph {et~al.}(2005)\citenamefont
  {Wallraff}, \citenamefont {Schuster}, \citenamefont {Blais}, \citenamefont
  {Frunzio}, \citenamefont {Majer}, \citenamefont {Devoret}, \citenamefont
  {Girvin},\ and\ \citenamefont {Schoelkopf}}]{wallraff2005approaching}%
  \BibitemOpen
  \bibfield  {author} {\bibinfo {author} {\bibfnamefont {A.}~\bibnamefont
  {Wallraff}}, \bibinfo {author} {\bibfnamefont {D.~I.}\ \bibnamefont
  {Schuster}}, \bibinfo {author} {\bibfnamefont {A.}~\bibnamefont {Blais}},
  \bibinfo {author} {\bibfnamefont {L.}~\bibnamefont {Frunzio}}, \bibinfo
  {author} {\bibfnamefont {J.}~\bibnamefont {Majer}}, \bibinfo {author}
  {\bibfnamefont {M.~H.}\ \bibnamefont {Devoret}}, \bibinfo {author}
  {\bibfnamefont {S.~M.}\ \bibnamefont {Girvin}}, \ and\ \bibinfo {author}
  {\bibfnamefont {R.~J.}\ \bibnamefont {Schoelkopf}},\ }\href@noop {}
  {\bibfield  {journal} {\bibinfo  {journal} {Physical review letters}\
  }\textbf {\bibinfo {volume} {95}},\ \bibinfo {pages} {060501} (\bibinfo
  {year} {2005})}\BibitemShut {NoStop}%
\bibitem [{\citenamefont {Gambetta}\ \emph {et~al.}(2007)\citenamefont
  {Gambetta}, \citenamefont {Braff}, \citenamefont {Wallraff}, \citenamefont
  {Girvin},\ and\ \citenamefont {Schoelkopf}}]{gambetta2007protocols}%
  \BibitemOpen
  \bibfield  {author} {\bibinfo {author} {\bibfnamefont {J.}~\bibnamefont
  {Gambetta}}, \bibinfo {author} {\bibfnamefont {W.~A.}\ \bibnamefont {Braff}},
  \bibinfo {author} {\bibfnamefont {A.}~\bibnamefont {Wallraff}}, \bibinfo
  {author} {\bibfnamefont {S.~M.}\ \bibnamefont {Girvin}}, \ and\ \bibinfo
  {author} {\bibfnamefont {R.~J.}\ \bibnamefont {Schoelkopf}},\ }\href@noop {}
  {\bibfield  {journal} {\bibinfo  {journal} {Physical Review A}\ }\textbf
  {\bibinfo {volume} {76}},\ \bibinfo {pages} {012325} (\bibinfo {year}
  {2007})}\BibitemShut {NoStop}%
\bibitem [{\citenamefont {Ben-Hur}\ \emph {et~al.}(2001)\citenamefont
  {Ben-Hur}, \citenamefont {Horn}, \citenamefont {Siegelmann},\ and\
  \citenamefont {Vapnik}}]{ben2001support}%
  \BibitemOpen
  \bibfield  {author} {\bibinfo {author} {\bibfnamefont {A.}~\bibnamefont
  {Ben-Hur}}, \bibinfo {author} {\bibfnamefont {D.}~\bibnamefont {Horn}},
  \bibinfo {author} {\bibfnamefont {H.~T.}\ \bibnamefont {Siegelmann}}, \ and\
  \bibinfo {author} {\bibfnamefont {V.}~\bibnamefont {Vapnik}},\ }\href@noop {}
  {\bibfield  {journal} {\bibinfo  {journal} {Journal of machine learning
  research}\ }\textbf {\bibinfo {volume} {2}},\ \bibinfo {pages} {125}
  (\bibinfo {year} {2001})}\BibitemShut {NoStop}%
\bibitem [{\citenamefont {Reagor}\ \emph {et~al.}(2018)\citenamefont {Reagor},
  \citenamefont {Osborn}, \citenamefont {Tezak}, \citenamefont {Staley},
  \citenamefont {Prawiroatmodjo}, \citenamefont {Scheer}, \citenamefont
  {Alidoust}, \citenamefont {Sete}, \citenamefont {Didier}, \citenamefont
  {da~Silva} \emph {et~al.}}]{reagor2018demonstration}%
  \BibitemOpen
  \bibfield  {author} {\bibinfo {author} {\bibfnamefont {M.}~\bibnamefont
  {Reagor}}, \bibinfo {author} {\bibfnamefont {C.~B.}\ \bibnamefont {Osborn}},
  \bibinfo {author} {\bibfnamefont {N.}~\bibnamefont {Tezak}}, \bibinfo
  {author} {\bibfnamefont {A.}~\bibnamefont {Staley}}, \bibinfo {author}
  {\bibfnamefont {G.}~\bibnamefont {Prawiroatmodjo}}, \bibinfo {author}
  {\bibfnamefont {M.}~\bibnamefont {Scheer}}, \bibinfo {author} {\bibfnamefont
  {N.}~\bibnamefont {Alidoust}}, \bibinfo {author} {\bibfnamefont {E.~A.}\
  \bibnamefont {Sete}}, \bibinfo {author} {\bibfnamefont {N.}~\bibnamefont
  {Didier}}, \bibinfo {author} {\bibfnamefont {M.~P.}\ \bibnamefont
  {da~Silva}},  \emph {et~al.},\ }\href@noop {} {\bibfield  {journal} {\bibinfo
   {journal} {Science advances}\ }\textbf {\bibinfo {volume} {4}},\ \bibinfo
  {pages} {eaao3603} (\bibinfo {year} {2018})}\BibitemShut {NoStop}%
\bibitem [{\citenamefont {Shannon}(1959)}]{shannon1959coding}%
  \BibitemOpen
  \bibfield  {author} {\bibinfo {author} {\bibfnamefont {C.~E.}\ \bibnamefont
  {Shannon}},\ }\href@noop {} {\bibfield  {journal} {\bibinfo  {journal} {IRE
  Nat. Conv. Rec}\ }\textbf {\bibinfo {volume} {4}},\ \bibinfo {pages} {1}
  (\bibinfo {year} {1959})}\BibitemShut {NoStop}%
\bibitem [{\citenamefont {Dr{\'e}au}\ \emph {et~al.}(2013)\citenamefont
  {Dr{\'e}au}, \citenamefont {Spinicelli}, \citenamefont {Maze}, \citenamefont
  {Roch},\ and\ \citenamefont {Jacques}}]{dreau2013single}%
  \BibitemOpen
  \bibfield  {author} {\bibinfo {author} {\bibfnamefont {A.}~\bibnamefont
  {Dr{\'e}au}}, \bibinfo {author} {\bibfnamefont {P.}~\bibnamefont
  {Spinicelli}}, \bibinfo {author} {\bibfnamefont {J.}~\bibnamefont {Maze}},
  \bibinfo {author} {\bibfnamefont {J.-F.}\ \bibnamefont {Roch}}, \ and\
  \bibinfo {author} {\bibfnamefont {V.}~\bibnamefont {Jacques}},\ }\href@noop
  {} {\bibfield  {journal} {\bibinfo  {journal} {Physical review letters}\
  }\textbf {\bibinfo {volume} {110}},\ \bibinfo {pages} {060502} (\bibinfo
  {year} {2013})}\BibitemShut {NoStop}%
\bibitem [{\citenamefont {Hann}\ \emph {et~al.}(2018)\citenamefont {Hann},
  \citenamefont {Elder}, \citenamefont {Wang}, \citenamefont {Chou},
  \citenamefont {Schoelkopf},\ and\ \citenamefont {Jiang}}]{hann2018robust}%
  \BibitemOpen
  \bibfield  {author} {\bibinfo {author} {\bibfnamefont {C.~T.}\ \bibnamefont
  {Hann}}, \bibinfo {author} {\bibfnamefont {S.~S.}\ \bibnamefont {Elder}},
  \bibinfo {author} {\bibfnamefont {C.~S.}\ \bibnamefont {Wang}}, \bibinfo
  {author} {\bibfnamefont {K.}~\bibnamefont {Chou}}, \bibinfo {author}
  {\bibfnamefont {R.~J.}\ \bibnamefont {Schoelkopf}}, \ and\ \bibinfo {author}
  {\bibfnamefont {L.}~\bibnamefont {Jiang}},\ }\href@noop {} {\bibfield
  {journal} {\bibinfo  {journal} {Physical Review A}\ }\textbf {\bibinfo
  {volume} {98}},\ \bibinfo {pages} {022305} (\bibinfo {year}
  {2018})}\BibitemShut {NoStop}%
\bibitem [{\citenamefont {Boissonneault}\ \emph {et~al.}(2009)\citenamefont
  {Boissonneault}, \citenamefont {Gambetta},\ and\ \citenamefont
  {Blais}}]{boissonneault2009dispersive}%
  \BibitemOpen
  \bibfield  {author} {\bibinfo {author} {\bibfnamefont {M.}~\bibnamefont
  {Boissonneault}}, \bibinfo {author} {\bibfnamefont {J.~M.}\ \bibnamefont
  {Gambetta}}, \ and\ \bibinfo {author} {\bibfnamefont {A.}~\bibnamefont
  {Blais}},\ }\href@noop {} {\bibfield  {journal} {\bibinfo  {journal}
  {Physical Review A}\ }\textbf {\bibinfo {volume} {79}},\ \bibinfo {pages}
  {013819} (\bibinfo {year} {2009})}\BibitemShut {NoStop}%
\bibitem [{\citenamefont {Baum}\ \emph {et~al.}(1970)\citenamefont {Baum},
  \citenamefont {Petrie}, \citenamefont {Soules},\ and\ \citenamefont
  {Weiss}}]{baum1970maximization}%
  \BibitemOpen
  \bibfield  {author} {\bibinfo {author} {\bibfnamefont {L.~E.}\ \bibnamefont
  {Baum}}, \bibinfo {author} {\bibfnamefont {T.}~\bibnamefont {Petrie}},
  \bibinfo {author} {\bibfnamefont {G.}~\bibnamefont {Soules}}, \ and\ \bibinfo
  {author} {\bibfnamefont {N.}~\bibnamefont {Weiss}},\ }\href@noop {}
  {\bibfield  {journal} {\bibinfo  {journal} {The annals of mathematical
  statistics}\ }\textbf {\bibinfo {volume} {41}},\ \bibinfo {pages} {164}
  (\bibinfo {year} {1970})}\BibitemShut {NoStop}%
\bibitem [{\citenamefont {Schreiber}(2017)}]{schreiber2017pomegranate}%
  \BibitemOpen
  \bibfield  {author} {\bibinfo {author} {\bibfnamefont {J.}~\bibnamefont
  {Schreiber}},\ }\href@noop {} {\bibfield  {journal} {\bibinfo  {journal} {The
  Journal of Machine Learning Research}\ }\textbf {\bibinfo {volume} {18}},\
  \bibinfo {pages} {5992} (\bibinfo {year} {2017})}\BibitemShut {NoStop}%
\bibitem [{\citenamefont {Picot}\ \emph {et~al.}(2008)\citenamefont {Picot},
  \citenamefont {Lupac{\c{s}}cu}, \citenamefont {Saito}, \citenamefont
  {Harmans},\ and\ \citenamefont {Mooij}}]{picot2008role}%
  \BibitemOpen
  \bibfield  {author} {\bibinfo {author} {\bibfnamefont {T.}~\bibnamefont
  {Picot}}, \bibinfo {author} {\bibfnamefont {A.}~\bibnamefont
  {Lupac{\c{s}}cu}}, \bibinfo {author} {\bibfnamefont {S.}~\bibnamefont
  {Saito}}, \bibinfo {author} {\bibfnamefont {C.~J. P.~M.}\ \bibnamefont
  {Harmans}}, \ and\ \bibinfo {author} {\bibfnamefont {J.~E.}\ \bibnamefont
  {Mooij}},\ }\href@noop {} {\bibfield  {journal} {\bibinfo  {journal}
  {Physical Review B}\ }\textbf {\bibinfo {volume} {78}},\ \bibinfo {pages}
  {132508} (\bibinfo {year} {2008})}\BibitemShut {NoStop}%
\bibitem [{\citenamefont {Schuster}\ \emph {et~al.}(2005)\citenamefont
  {Schuster}, \citenamefont {Wallraff}, \citenamefont {Blais}, \citenamefont
  {Frunzio}, \citenamefont {Huang}, \citenamefont {Majer}, \citenamefont
  {Girvin},\ and\ \citenamefont {Schoelkopf}}]{schuster2005ac}%
  \BibitemOpen
  \bibfield  {author} {\bibinfo {author} {\bibfnamefont {D.~I.}\ \bibnamefont
  {Schuster}}, \bibinfo {author} {\bibfnamefont {A.}~\bibnamefont {Wallraff}},
  \bibinfo {author} {\bibfnamefont {A.}~\bibnamefont {Blais}}, \bibinfo
  {author} {\bibfnamefont {L.}~\bibnamefont {Frunzio}}, \bibinfo {author}
  {\bibfnamefont {R.-S.}\ \bibnamefont {Huang}}, \bibinfo {author}
  {\bibfnamefont {J.}~\bibnamefont {Majer}}, \bibinfo {author} {\bibfnamefont
  {S.~M.}\ \bibnamefont {Girvin}}, \ and\ \bibinfo {author} {\bibfnamefont
  {R.~J.}\ \bibnamefont {Schoelkopf}},\ }\href@noop {} {\bibfield  {journal}
  {\bibinfo  {journal} {Physical Review Letters}\ }\textbf {\bibinfo {volume}
  {94}},\ \bibinfo {pages} {123602} (\bibinfo {year} {2005})}\BibitemShut
  {NoStop}%
\bibitem [{\citenamefont {Jin}\ \emph {et~al.}(2015)\citenamefont {Jin},
  \citenamefont {Kamal}, \citenamefont {Sears}, \citenamefont {Gudmundsen},
  \citenamefont {Hover}, \citenamefont {Miloshi}, \citenamefont {Slattery},
  \citenamefont {Yan}, \citenamefont {Yoder}, \citenamefont {Orlando} \emph
  {et~al.}}]{jin2015thermal}%
  \BibitemOpen
  \bibfield  {author} {\bibinfo {author} {\bibfnamefont {X.}~\bibnamefont
  {Jin}}, \bibinfo {author} {\bibfnamefont {A.}~\bibnamefont {Kamal}}, \bibinfo
  {author} {\bibfnamefont {A.}~\bibnamefont {Sears}}, \bibinfo {author}
  {\bibfnamefont {T.}~\bibnamefont {Gudmundsen}}, \bibinfo {author}
  {\bibfnamefont {D.}~\bibnamefont {Hover}}, \bibinfo {author} {\bibfnamefont
  {J.}~\bibnamefont {Miloshi}}, \bibinfo {author} {\bibfnamefont
  {R.}~\bibnamefont {Slattery}}, \bibinfo {author} {\bibfnamefont
  {F.}~\bibnamefont {Yan}}, \bibinfo {author} {\bibfnamefont {J.}~\bibnamefont
  {Yoder}}, \bibinfo {author} {\bibfnamefont {T.}~\bibnamefont {Orlando}},
  \emph {et~al.},\ }\href@noop {} {\bibfield  {journal} {\bibinfo  {journal}
  {Physical Review Letters}\ }\textbf {\bibinfo {volume} {114}},\ \bibinfo
  {pages} {240501} (\bibinfo {year} {2015})}\BibitemShut {NoStop}%
\end{thebibliography}
\end{document}